\documentclass[10pt,preprint,aps]{revtex4-1}
\usepackage[latin9]{inputenc}
\usepackage{xcolor}
\usepackage{pdfcolmk}
\usepackage{array}
\usepackage{float}
\usepackage{rotfloat}
\usepackage{multirow}
\usepackage{amsmath}
\usepackage{amssymb}
\usepackage{graphicx}
\PassOptionsToPackage{normalem}{ulem}
\usepackage{ulem}
\usepackage{subscript}
\usepackage[unicode=true,
 bookmarks=true,bookmarksnumbered=false,bookmarksopen=false,
 breaklinks=false,pdfborder={0 0 1},backref=false,colorlinks=false]
 {hyperref}
\hypersetup{pdftitle={quantum state preparation}}

\makeatletter

\providecommand{\tabularnewline}{\\}
\providecolor{lyxadded}{rgb}{0,0,1}
\providecolor{lyxdeleted}{rgb}{1,0,0}

\newenvironment{lyxlist}[1]
{\begin{list}{}
{\settowidth{\labelwidth}{#1}
 \setlength{\leftmargin}{\labelwidth}
 \addtolength{\leftmargin}{\labelsep}
 }}
{\end{list}}

\makeatother

\begin{document}

\title{\textcolor{black}{Canceling spin-dependent contributions and systematic
shifts in precision spectroscopy of the molecular hydrogen ions }}

\author{S. Schiller}

\address{Institut für Experimentalphysik, Heinrich-Heine-Universität Düsseldorf,
40225 Düsseldorf, Germany}

\author{V.~I. Korobov}

\address{Bogoliubov Laboratory of Theoretical Physics, Joint Institute for
Nuclear Research, 141980 Dubna, Russia}
\begin{abstract}
We consider the application of a basic principle of quantum theory,
the tracelessness of a certain class of hamiltonians, to the precision
spectroscopy of the molecular hydrogen ions. We show that it is possible
to obtain the spin-averaged transition frequencies between states
from a simple weighted sum of experimentally accessible spin-dependent
transition frequencies. We discuss the cases ${\rm H}_{2}^{+}$ and
${\rm HD}^{+}$, which are distinct in the multipole character of
their rovibrational transitions. Inclusion of additional frequencies
permits canceling also the electric quadrupole shift, the Zeeman shift
and partially the Stark shift. In this context, we find that measuring
electric quadrupole transitions in ${\rm HD}^{+}$ is advantageous.
The required experimental effort appears reasonable. 
\end{abstract}
\maketitle
The precision spectroscopy of isolated molecules is making strong
advances thanks to novel techniques of trapping, cooling and manipulation.
One family of molecules, the molecular hydrogen ions, is particularly
attractive because their transition frequencies can be calculated
\textit{ab initio} with a precision that challenges current experimental
approaches \cite{Korobov2017a}.

The comparison of theoretical with experimental frequencies allows
extracting the values of certain fundamental constants, such as the
electron-to-proton mass ratio and the Rydberg constant \cite{Koelemeij2007,Bressel2012,Biesheuvel2016,Alighanbari2018}.
In this endeavor, one is faced with the problem that the theoretical
frequencies are not just given by the solution of the Schrödinger
three-body problem, but include relativistic and QED contributions,
as well as spin-dependent contributions. In addition, the experimentally
measured frequencies are perturbed by external fields whose strength
cannot always be determined accurately enough.

The spin-dependent contributions can be calculated and, in fact, the
accuracy of the calculation has made impressive progress in recent
years \cite{Korobov2016}. Nevertheless, it will become harder to
push the accuracy further. Thus, it is worthwhile to consider whether
there are experimentally viable approaches to determine the (not directly
observable) spin-averaged transition frequency from a \textit{combination}
of measured transition frequencies. The solution to this query makes
use of the mathematical property of a class of hamiltonians, the tracelessness.
In atomic physics, this approach is well-known; here the computed
``center of gravity'' of the fine-structure or of the hyperfine
structure of a level is often considered. In the field of metrology
it is for example used to characterize the helium isotope shift \cite{Pachucki2015}. 

We further consider an important subset of systematic shifts affecting
the transition frequencies. When the shifts are small, so that first-order
and second-order perturbation is applicable, their mathematical structure
is such that an appropriate average over them vanishes exactly. This
principle was recently implicitly used by Karr et al. \cite{Karr2016}
in an analysis of promising transitions for the precision spectroscopy
of ${\rm H}_{2}^{+}$. Here we illuminate the principle from a broader
perspective and show the close relationship to the concept of the
spin-averaged transition frequency determination.

Composite frequencies being at the focus of the present work, it complements
our previous discussion \cite{Schiller2014}, in which we considered
the combination of different rovibrational transitions for the cancellation
of systematic shifts. Here, individual rovibrational transitions are
considered.

The paper is structured as follows. In Sec.~I we derive cancellation
conditions from the tracelessness of the hamiltonian. In Sec.~II
it is shown that they lead to the possibility of determining the spin-averaged
frequency via the combination of a set of electric-dipole transitions.
In Sec.~III the conditions for cancellation of three types of systematic
shifts are derived: linear Zeeman shift, quadratic Zeeman shift, and
electric quadrupole shift. A fourth systematic shift, the Stark shift,
can be partially canceled, since its tensor part is canceled whenever
the quadrupole shift is. In Sec.~IV we state how spin contributions
and shifts can be simultaneously canceled. In Sec.~V we apply the
general discussion of spin structure cancellation to the molecular
hydrogen ions ${\rm H}_{2}^{+}$ and ${\rm HD}^{+}$. In Sec.~VI
we show that an obvious extension leads to cancellation of the linear
Zeeman shift as well. Further shifts can also be canceled by including
more transitions. Subsequently, the general method from Sec.~IV is
applied. For each approach, the systematic shifts stemming from the
un-cancelled perturbations are evaluated and discussed, leading to
estimates for the achievable residual shifts. In Sec.~VII we discuss
a more efficient approach for canceling both spin-structure and systematic
shifts. Sec.~VIII draws the conclusions.

\section{Cancellations: Elementary Considerations}

Consider a complete set $\{|\psi_{i}\rangle\}$ of basis states, where
$i$ denotes the set of quantum numbers uniquely identifying a state,
and a generic traceless operator $H^{1}$. The traceless property
implies, 

\begin{equation}
0={\rm Tr}(H^{1})=\sum_{i}\langle\psi_{i}|H^{1}|\psi_{i}\rangle\,.
\end{equation}

The trace can be evaluated in any basis. In particular, we can choose
a set of basis states that diagonalizes the operator, its eigenstates. 

Consider now the operator $H^{1}$ to be a contribution to the total
hamiltonian $H$. The tracelessness of $H^{1}$ then yields a relationship
between the energy contributions $E_{\xi}^{1}$:

\begin{equation}
0={\rm Tr}(H^{1})=\sum_{\xi}d_{\xi}E_{\xi}^{1}\,,\label{eq:tracelessness for energies}
\end{equation}
where $d_{\xi}$ is the degeneracy factor for those states having
the same energy contribution $E_{\xi}^{1}$, and $\xi$ denotes the
set of quantum numbers identifying the different \textit{energy levels}.
The expression implies the existence of positive and negative energy
contributions.

Often, one is concerned with states that are characterized by a total
angular momentum quantum number, $J$, and a magnetic quantum number,
$J_{z}$. We may choose these states as basis states. In case of rotational
invariance of the hamiltonian $H^{1}$, its expectation values are
independent of $J_{z}$. In this case the above expression takes the
form

\begin{equation}
0={\rm Tr}(H^{1})=\sum_{n,J}(2J+1)\langle\psi_{n,J}|H^{1}|\psi_{n,J}\rangle\,,
\end{equation}
where $2J+1$ is the magnetic degeneracy factor, and $n$ denotes
all other quantum numbers necessary to identify the state.

The typical situation in atomic and molecular physics is that the
system is described by a total hamiltonian which is the sum of a ``dominant''
term, $H^{0}$ (spin-independent or spin-averaged) and several traceless
perturbations, each denoted by $H^{j}$. The energy of an eigenstate
$|\psi_{k}\rangle$ of the total hamiltonian is approximated by first-order
perturbation theory,

\begin{align}
E_{k} & =E_{k}^{0}+\sum_{j}E_{k}^{j}\\
 & =\langle\psi_{k}^{0}|H^{0}|\psi_{k}^{0}\rangle+\sum_{j}\langle\psi_{k}^{0}|H^{j}|\psi_{k}^{0}\rangle\,.\nonumber 
\end{align}
The expectation values are taken for the unperturbed states $|\psi_{k}^{0}\rangle$,
the eigenstates of the dominant hamiltonian $H^{0}$. In fact, the
quantum numbers for a state $k$ can be written more explicitly as
$p,\,m$, where $p$ refers to the set describing the state space
of the dominant hamiltonian, and $m$ to the space of the perturbation
hamiltonian. The trace condition is then

\begin{equation}
0={\rm Tr}(H^{j})=\sum_{m}\langle\psi_{p,m}^{0}|H^{j}|\psi_{p,m}^{0}\rangle\,.
\end{equation}

When we perform spectroscopy, we measure energy differences $E_{k'}-E_{k}=E_{p',m'}-E_{p,m}$.
The question we pose is whether a combination of such energy differences,
provided they are experimentally accessible, allows to determine the
dominant contribution $E_{k'}^{0}-E_{k}^{0}$, which may be of interest.
To this end, we consider a linear combination of transition frequencies,
the ``traceless'' frequency $f_{{\rm t}}$, of the form

\begin{equation}
h\,f_{{\rm t}}(p\rightarrow p')=\sum_{m,m'}\alpha(p',m';p,m)(E_{p',m'}-E_{p,m})\,,\label{eq:f_c not simplified}
\end{equation}
with (positive or negative) weights $\alpha$. This ansatz is successful,
if the three conditions hold:
\begin{lyxlist}{00.00.0000}
\item [{(I)}] all transition frequencies $h\,f(k\rightarrow k')=E_{p',m'}-E_{p,m}$
are experimentally measurable (necessarily, the transitions must be
allowed by selection rules), 
\item [{(II)}] $\sum_{m'}\alpha(p',m';p,m)=1/{\cal N}$ is satisfied
for all possible values $m$, 
\item [{(III)}] $\sum_{m}\alpha(p',m';p,m)=1/{\cal N}'$ is satisfied for
all possible values $m'$.
\end{lyxlist}
Here ${\cal N}=\sum_{m}1$ is the total number of states for given
$p$, etc. (II) and (III) simply mean that in the set of selected
transition frequencies each initial state should occur with equal
total weight $1/{\cal N}$ and each final state should occur with
equal total weight $1/{\cal N}'$ .

Then, eq.~(\ref{eq:f_c not simplified}) simplifies to
\begin{align}
h\,f_{{\rm t}}(p\rightarrow p') & =\sum_{m,m'}\alpha(p',m';p,m)E_{p',m'}-\sum_{m,m'}\alpha(p',m';p,m)E_{p,m}\label{eq:traceless frequency - first definition}\\
 & =\sum_{m'}(1/{\cal N}')(E_{p'}^{0}+\sum_{j}E_{p',m'}^{j})-\sum_{m}(1/{\cal N})(E_{p}^{0}+\sum_{j}E_{p,m}^{j})\nonumber \\
 & =\sum_{m'}(1/{\cal N}')E_{p'}^{0}-\sum_{m}(1/{\cal N})E_{p}^{0}\nonumber \\
 & =E_{p'}^{0}-E_{p}^{0}\,,\nonumber 
\end{align}
where we have made use of the tracelessness, $0=\sum_{m'}E_{p',m'}^{j}$
$=\sum_{m}E_{p,m}^{j}$.

\section{Cancellation of Spin Structure Contributions\label{sec:Cancellation-of-Spin Structure Contributions}}

Generally, the effective spin hamiltonian, comprising fine-structure,
hyperfine structure, and interaction of the electron with the nuclear
quadrupole moment, may be expressed as a sum of traceless operators,
\begin{equation}
H^{{\rm spin}}=\sum_{j}{\cal E}_{j}(T_{a}^{(k)}\cdot U_{b}^{(k)})\label{eq:HFS hamiltonian}
\end{equation}
where $T_{a}^{(k)}$ and $U_{b}^{(k)}$ are some irreducible tensors
of spin or orbital operators and dot is a tensor scalar product 
\[
T_{a}^{(k)}\cdot U_{b}^{(k)}=\sum_{\mu}\,(-1)^{\mu}\,T_{a,\mu}^{(k)}U_{b,-\mu}^{(k)}\,.
\]
A first example is the spin-orbit interaction ${\cal E}_{j}(\mathbf{L}\cdot\mathbf{s}_{j})$,
where $\mathbf{L}$ is the operator of the total orbital angular momentum
and $\mathbf{s}_{j}$ is the spin operator of particle $j$. A second
example is the tensor quadrupole interaction,
\[
{\cal E}_{j}\left\{ \mathbf{L}^{2}(\mathbf{s}_{k}\cdot\mathbf{s}_{l})-3\left[(\mathbf{L}\cdot\mathbf{s}_{k})(\mathbf{L}\cdot\mathbf{s}_{l})+(\mathbf{L}\cdot\mathbf{s}_{l})(\mathbf{L}\cdot\mathbf{s}_{k})\right]\right\} \,.
\]
This form is provided by the spin-dependent part of the Breit-Pauli
hamiltonian. Higher-order corrections enter either as corrections
to the coefficients of the already existing interactions, or as new,
more complicated irreducible tensor (traceless) interactions, or they
contribute to the spin-averaged part of the energy of a state.

If the system is composed by three particles with spin, we introduce
the two-particle spin operator ${\bf F}={\bf s}_{1}+{\bf s}_{2}$,
and the three-particle spin operator ${\bf S}={\rm {\bf F}}+{\bf s}_{3}$,
which generally do not commute with the total hamiltonian. The total
angular momentum, $\mathbf{J}=\mathbf{S}+\mathbf{N}$, does commute
with $H$, and $J,\,J_{z}$ are good quantum numbers. As basis states
we can take pure angular momentum states or the eigenstates of the
total hamiltonian $H^{0}+H^{{\rm spin}}$. In both cases we can denote
them by $|p,FSJJ_{z}\rangle$. In the latter case, the numbers $F,\,S$
are chosen as the integers closest to the numbers $\bar{F},\,\bar{S}$
resulting from the expectation values $\langle{\bf F}^{2}\rangle=\hbar^{2}\bar{F}(\bar{F}+1)$
and $\langle{\bf S}^{2}\rangle=\hbar^{2}\bar{S}(\bar{S}+1)$ in the
given state. In absence of magnetic field, the eigenstates are degenerate
in $J_{z}$. We denote by $E_{p,FSJ}^{{\rm spin}}$ the perturbation
energy of state $|p,FSJJ_{z}\rangle$. 

The following sum rule holds for the traceless spin hamiltonian:

\begin{equation}
\sum_{FSJ}(2J+1)\,E_{p,FSJ}^{{\rm spin}}=0\,.\label{eq:tracelessness of spin hamiltonian}
\end{equation}
The sum is over all spin states. The proof is straightforward because
it is fulfilled for each individual term in eq.~(\ref{eq:HFS hamiltonian})
\cite{Schwarz1955}.

We now consider transitions and show how the traceless frequency eq.~(\ref{eq:f_c not simplified})
can be implemented. 

A transition frequency between two particular spin states of two rovibrational
levels is
\begin{equation}
h\,f(pFSJ\rightarrow p'F'S'J')=E_{p'}^{0}-E_{p}^{0}+E_{p',F'S'J'}^{{\rm spin}}-E_{p,FSJ}^{{\rm spin}}\,.
\end{equation}
In E1-allowed transitions, $F=F',\,S=S'$ do not change, but the values
must nevertheless be included to uniquely identify the state. 

The following quantity vanishes, 
\begin{equation}
{\normalcolor \sum_{(FSJ)\rightarrow(FSJ')}}(2{\normalcolor J'}+1)(E_{p',FSJ'}^{{\rm spin}}-E_{p,FSJ}^{{\rm spin}})=0\,,\label{eq:rule_trans}
\end{equation}
where the sum is over all E1-allowed transitions between levels $p$
and $p'$. 

The proof is as follows. The E1 selection rule is $J\to J'=J,\,J\pm1$.
If $J=0$, then $J'=1$.

We use the obvious relation 
\[
\sum_{J'=J,J\pm1}(2J'+1)=(2(J+1)+1)+(2J+1)+(2(J-1)+1)=3(2J+1),
\]
or for $J=0$: $3(2J+1)=3=(2J'+1)$. That proves that the l.h.s.~of
eq.~(\ref{eq:rule_trans}) may be written as 
\[
\sum_{(FSJ')}(2J'+1)E_{p',FSJ'}^{{\rm spin}}-3\sum_{(FSJ)}(2J+1)\,E_{p,FSJ}^{{\rm spin}}\,,
\]
which vanishes on account of tracelessness, eq.~(\ref{eq:tracelessness of spin hamiltonian}).
Thus, the traceless frequency constructed from all E1-allowed transition
frequencies 

\begin{align}
f_{{\rm t}}^{{\rm spin}} & ={\cal M}^{-1}\sum_{(FSJ)\rightarrow(FSJ')}(2J'+1)f(pFSJ\rightarrow p'FSJ')\\
 & =(E_{p'}^{0}-E_{p}^{0})/h
\end{align}
is equal to the spin-averaged transition frequency $f^{0}(p\rightarrow p')$.
Here, ${\cal M=}\sum_{(FSJ)\rightarrow(FSJ')}(2J'+1)$ is a normalization
factor. In Sec.~\ref{subsec:The-molecular-ion HDplus} below, we
show an example of the set of transitions that are included in the
sum. 

When E1 transitions do not exist, the treatment must be modified,
see Sec.~\ref{subsec:The-molecular-ion H2plus} below.

\section{Cancellation of Systematic Shifts\label{sec:Application-to-the cancellation of systematic shifts}}

\subsection{Linear Zeeman shift and electric quadrupole shift: a simple model}

We now consider a simple example: a quantum system possessing angular
momentum ${\bf J}$ and an electric quadrupole moment, exposed to
an external magnetic field ${\bf B}$ pointing along the $z$-axis
and to an external electric field gradient $V_{zz}$. Assume the magnetic
and electric quadrupole perturbation hamiltonian to be of the form

\begin{equation}
H_{p}^{1}=-\mu_{B}g(p){\bf J\cdot B}+d(p)V_{zz}({\bf J^{2}}-3J_{z}^{2})\,.\label{eq:simple model of ZS and EQS}
\end{equation}
Both contributions are traceless. The coefficients $g$ and $d$ may
in general differ in different states $p$ of the unperturbed hamiltonian
$H^{0}$. The state quantum numbers summarized by $p$ include $J$,
which is assumed fixed in this example. The angular momentum projection
quantum number $J_{z}$ now plays the role of the quantum number $m$.
The linear Zeeman (LZ) shift and the electric quadrupole (EQ) shift
are computed in 1\textsuperscript{st}-order perturbation theory.
The tracelessness of $H_{p}^{1}$ is expressed by the sum rule,

\begin{equation}
\sum_{J_{z}}E_{p,J_{z}}^{1}=\sum_{J_{z}}-\mu_{B}g(p)J_{z}|{\bf B}|+d(p)V_{zz}(J(J+1)-3J_{z}^{2})=0,
\end{equation}
which holds independently of the specific values of $J$ (integer
or half-integer), $g$, $d$ (and thus of the specific level $p$)
and of the field strengths $|{\bf B}|$, and $V_{zz}$.

The following examples illustrate how to include this sum rule into
the traceless frequency. In Fig.~\ref{fig:magnetic and el.quadrupole transition diagram}~(a-c)
we show schematically some energy levels of the total hamiltonian,
a lower level $p$ having $J=0,\,1,$ or 2, and an upper level $p'$
having $J'=1,\,2,$ or 3. The energy of a state is $E_{p}^{0}+E_{p,J_{z}}^{1}$.
A number of electric dipole (E1) transitions between the two levels,
obeying the selection rules $\Delta J=J'-J=0,\,\pm1$, $\Delta J_{z}=0,\,\pm1$,
have been selected and assigned specific weights $\alpha(J',J_{z}';J,J_{z})$
satisfying the conditions (II) and (III). These weights can easily
be found by solving appropriate conditions. In the case Fig.~\ref{fig:magnetic and el.quadrupole transition diagram}~(b),
the 9 transitions $J_{z}\rightarrow J_{z}'$, which altogether address
every state of the lower level with the same weight (1/3) and every
state of the upper level with the same weight (1/5), serve to null
the effect of four interactions, proportional to $g(p,J)|{\bf B}|$,
$g(p',J')|{\bf B}|$, $d(p,J)V_{zz}$ and $d(p',J')V_{zz}$,

\[
h\,f_{{\rm t}}^{1}=h\sum_{J_{z},J_{z}'}\alpha(J',J_{z}';J,J_{z})\,f(p,\,J_{z}\rightarrow p',\,J_{z}')=E_{p'}^{0}-E_{p}^{0}\,.
\]
The specific cases (a-c) can easily be generalized to higher values
of $J$ and $J'=J+1$. 

The panels (d-e) in the same figure show cases of half-integer angular
momenta, $J=1/2$, $J'=3/2,\,5/2$, and electric-quadrupole (E2) transitions.
A suitable set of transitions, satisfying the selection rule for E2
transitions, and appropriate weights are found by inspection. For
example, panel (d) shows the case where the measurement of a total
of 4 individual transitions $f(p',\,J_{z}',\,p,\,J_{z})$ allows canceling
the effect of Zeeman and EQ interactions in both lower $(J=1/2)$
and upper $(J'=3/2)$ level, defined by four interactions. In panel
(e) $J'$ is larger, and now 6 transitions are required for canceling
again four interactions. However, in these particular cases having
$J=1/2$, the EQ interaction in the lower level vanishes, so effectively
only three interactions are canceled.

The cases $J\rightarrow J'=J$ are trivial, therefore only one example
is shown, panel (a), right-hand side.

We emphasize that the presented scheme for nulling the effects of
the Zeeman and quadrupole shifts is not unique. Other types of combinations
of transition frequencies are possible. Specifically for ${\rm H_{2}^{+}}$,
Karr et al. \cite{Karr2016} have considered a combination of frequencies
involving two different angular momentum sub-spaces in one of the
two levels (see also below). Such other combinations can be more efficient,
requiring a smaller number of transitions to be measured.

\subsection{Zeeman shift and electric quadrupole shift: realistic case}

The EQ interaction between an external field gradient and the electronic
plus nuclear charge distribution has the form
\[
H^{{\rm EQ'}}(p)\propto({\bf L\otimes L})^{(2)}\,,
\]
where the r.h.s.~is an irreducible tensor operator of rank 2. This
hamiltonian is traceless. Its effects are evaluated in first-order
perturbation theory, due to the smallness of the field gradient occurring
in experiments. The hamiltonian can be replaced by \cite{Bakalov2014includingcorrigendum}
\[
H^{{\rm EQ}}(p)\propto{\bf J}^{2}-3\hat{J_{z}}^{2}\,.
\]
The first-order energy shifts of a given state $(p,\,FSJ)$ are $E_{p,FSJ}^{{\rm EQ}}\propto J(J+1)-3J_{z}^{2}$,
the same as for the simple model above. 

The Stark shift arises from light fields and trap fields. In general,
it has a scalar, vector, and tensor contribution \cite{Schiller2014a}
(see eq.~(5) in ref.~\cite{Bakalov2015} for the formal expression
for the static Stark shift). The tensor contribution has the same
dependence on the angular momentum quantum numbers as the EQ shift.
Therefore cancellation of EQ shift implies cancellation of the tensor
Stark shift. 

In real systems, the interaction with the external magnetic field
is not of the form eq.~(\ref{eq:simple model of ZS and EQS}). Instead,
in general,

\[
H^{{\rm mag}}(p)=-\mu_{B}\left(g_{1}(p)\hat{s}_{1,z}+g_{2}(p)\hat{s}_{2,z}+g_{3}(p)\hat{s}_{3,z}+g_{L}(p)\hat{L}_{z}\right)B_{z}\,.
\]

This hamiltonian is traceless, and furthermore commutes with $\hat{J_{z}}$.
We can take advantage of the structure of the magnetic shifts that
occur in first order and in second order in $B_{z}$ and incorporate
them into the traceless frequency.

In first-order perturbation theory, 
\[
E_{p,FSJJ_{z}}^{{\rm LZ}}=\langle p,FSJJ_{z}|H^{{\rm mag}}(p)|p,FSJJ_{z}\rangle\propto J_{z}\,.
\]
The LZ shift is proportional to $J_{z}$ by virtue of the Wigner-Eckhart
theorem. 

At today's desired precision levels, it is insufficient to consider
only the LZ shift. It is necessary to also consider the quadratic
Zeeman (QZ) shift, which according to second-order perturbation theory
is

\begin{equation}
E_{p,FSJJ_{z}}^{{\rm QZ}}=\sum_{F'S'J'\ne FSJ}\frac{|\langle p,F'S'J'J_{z}|H^{{\rm mag}}(p)|p,FSJJ_{z}\rangle|^{2}}{E_{p,FSJ}^{0}-E_{p,F'S'J'}^{0}}\,.
\end{equation}
for a state state $|p,FSJJ_{z}\rangle$. The sum goes over all spin-structure
states $(F'S'J')$ but is limited to states in the same level $p$.
Also, $J_{z}'=J_{z}$ since $H^{{\rm mag}}$ commutes with the operator
$\hat{J_{z}}$. Because the denominator is anti-symmetric under ``state
exchange'' $(FSJ)\leftrightarrow(F'S'J')$, while the numerator is
symmetric, it follows that for any given $J_{z}$

\begin{equation}
\sum_{FSJ}E_{p,FSJJ_{z}}^{{\rm QZ}}=0\,.\label{eq:Nulling of quadratic ZS}
\end{equation}
 \textit{The sum over the quadratic Zeeman shifts of all spin states
in a given rovibrational level $p$ and having a given $J_{z}$ is
zero.} This sum rule does not contain any degeneracy factor since
$J_{z}$ is fixed. 

Note that there exists only one state having $J_{z}=J_{{\rm max}}=F+S+N$
and one having $J_{z}=-J_{{\rm max}}$ (stretched states). These states
therefore do not exhibit a QZ shift.

\section{Combining spin structure cancellation with systematic shift cancellation}

The standard situation in the description of the molecular hydrogen
ions is to consider the spin structure contributions and systematic
shifts for each rovibrational level $p$, independently of the others.
This is a good approximation because both types of contributions are
very small compared to the energy difference to neighboring rotational
levels of the same vibrational level and even smaller compared to
the energy difference to neighboring vibrational levels. The hamiltonians
$H^{j}$ of the perturbations therefore are effective hamiltonians,
i.e. they contain parameters that depend on the concrete level $p$:
$H^{j}=H^{j}(p)$. Given an arbitrary basis of spin states $q$ for
the particular level $p$, one can set up the hamiltonian matrix 
\begin{equation}
H_{q',q}^{{\rm pert}}(p)=\langle q'|H^{{\rm spin}}(p)+H^{{\rm mag}}(p)+H^{{\rm EQ}}(p)|q\rangle\,,
\end{equation}
and diagonalize it in order to find the eigenstates $m$ and the eigenenergies
$E_{p,m}^{{\rm pert}}$. Because each contribution in $H^{{\rm pert}}(p)$
is traceless, we have the sum rule
\begin{equation}
\sum_{q}H_{q,q}^{{\rm pert}}(p)=\sum_{m}E_{p,m}^{{\rm pert}}=0\,.
\end{equation}

We emphasize that this sum rule refers to the ``exact'' total perturbation
shifts, to \emph{all orders in any perturbation parameter}, e.g. the
magnetic field strength or electric field gradient strength. 

Since it is permissible to consider various orders of a particular
perturbation, e.g. the linear Zeeman shift and the quadratic Zeeman
shift, there necessarily follow separate sum rules for these orders.
Some of these have been already given above. Indeed, for practical
reasons it is useful to take this point of view: in precision spectroscopy,
one strives experimentally to make as many of the perturbations as
small as possible, so that the dominant contribution to $H^{{\rm pert}}$
is the spin structure $H^{{\rm spin}}$. The other perturbations are
then treated in first-order perturbation theory with respect to the
eigenstates of $H^{{\rm spin}}$, and only occasionally also second-order
perturbation theory is applied. The advantage of this approach is
that the perturbation energy of a state $(p,\,m)$ can then be written
as a \emph{sum of contributions}, and the total energy is
\[
E_{p,m}=E_{p}^{0}+E_{p,m}^{{\rm spin}}+E_{p,m}^{{\rm LZ}}+E_{p,m}^{{\rm QZ}}+E_{p,m}^{{\rm EQ}}\,.
\]
Here, $m$ corresponds to $FSJJ_{z}$. The spin interaction is treated
exactly (by diagonalization), while the Zeeman interaction is treated
to second-order perturbation theory (LZ, QZ), and the EQ interaction
only to first-order perturbation theory. All perturbations come with
their own sum rules, which have been presented above. They do differ
in form: 
\begin{itemize}
\item the LZS can be nulled, for any particular $FSJ$, by summing over
all $J_{z}$, or pairwise over $J_{z}$ and $-J_{z}$. 
\item the EQS can be nulled, for any particular $FSJ$, by summing over
all $J_{z}$. 
\item the QZ shift can be nulled, for any fixed $J_{z}$, by summing over
all states $FSJ$ containing this Zeeman state. 
\item The spin structure contributions can be nulled by summing over all
states $FSJJ_{z}$, or over all states $FSJ$ with $J_{z}=0$. 
\end{itemize}
Although these sum rules differ, not surprisingly they can nevertheless
be incorporated \textit{together} in a traceless frequency as in eq.~(\ref{eq:f_c not simplified}),
by defining 

\begin{equation}
h\,f_{{\rm t}}^{{\rm pert-free}}(p\rightarrow p')=\sum_{m,m'}'\tilde{\alpha}(p',m';p,m)(E_{p',m'}-E_{p,m})\,,\label{eq:f_t,tot}
\end{equation}
A sum over all possible transitions $m\rightarrow m'$ is taken, with
the following restrictions:

(1) Only those $FSJ\rightarrow F'S'J'$ transitions are included that
allow the spin structure cancellation as in the case of absence of
external fields. Depending on the particular transition $p\rightarrow p'$,
either each upper state $F'S'J'$ is associated with one lower state
$FSJ$ only, or vice-versa (for the whole set of lower states). Here
one disregards the concrete Zeeman components. 

\begin{figure}[H]
\begin{centering}
\includegraphics[width=9cm]{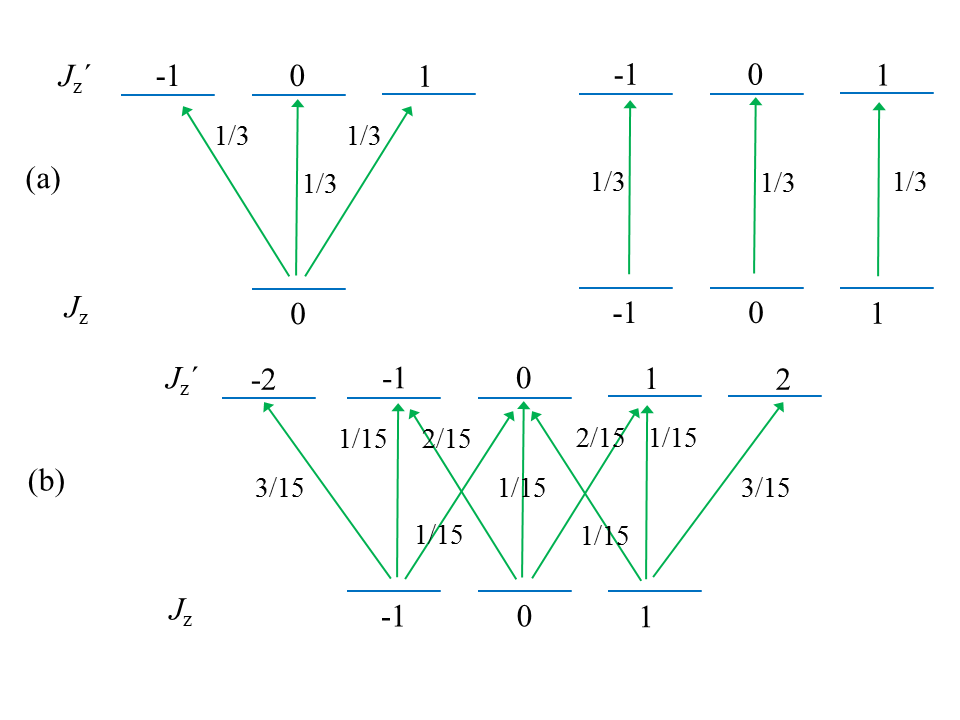}\includegraphics[width=0.5\columnwidth]{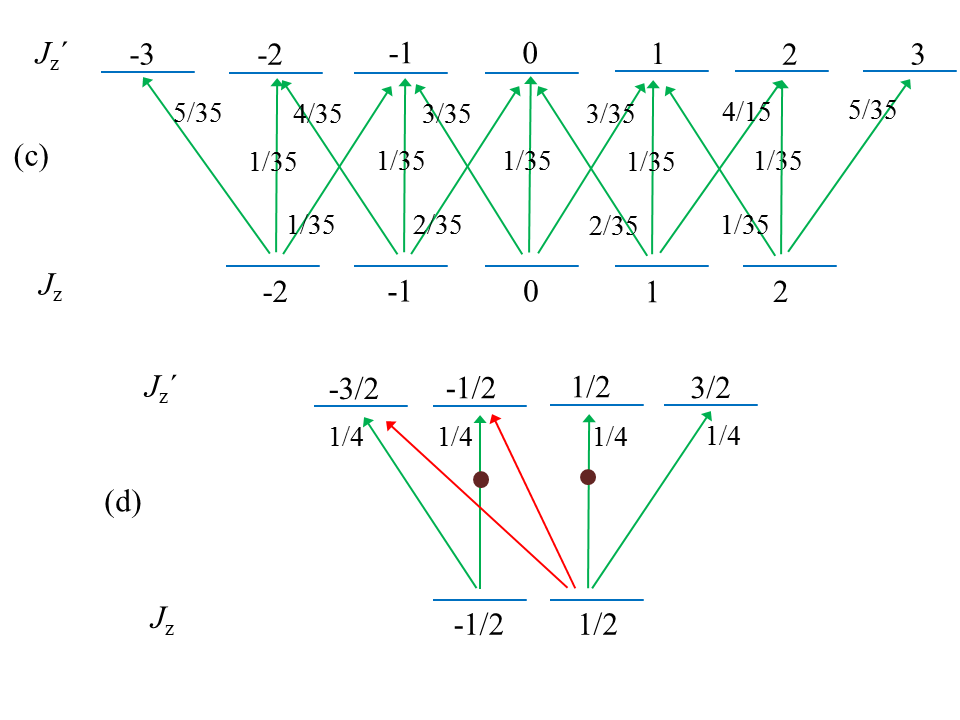}
\par\end{centering}
\begin{centering}
\includegraphics[width=0.5\columnwidth]{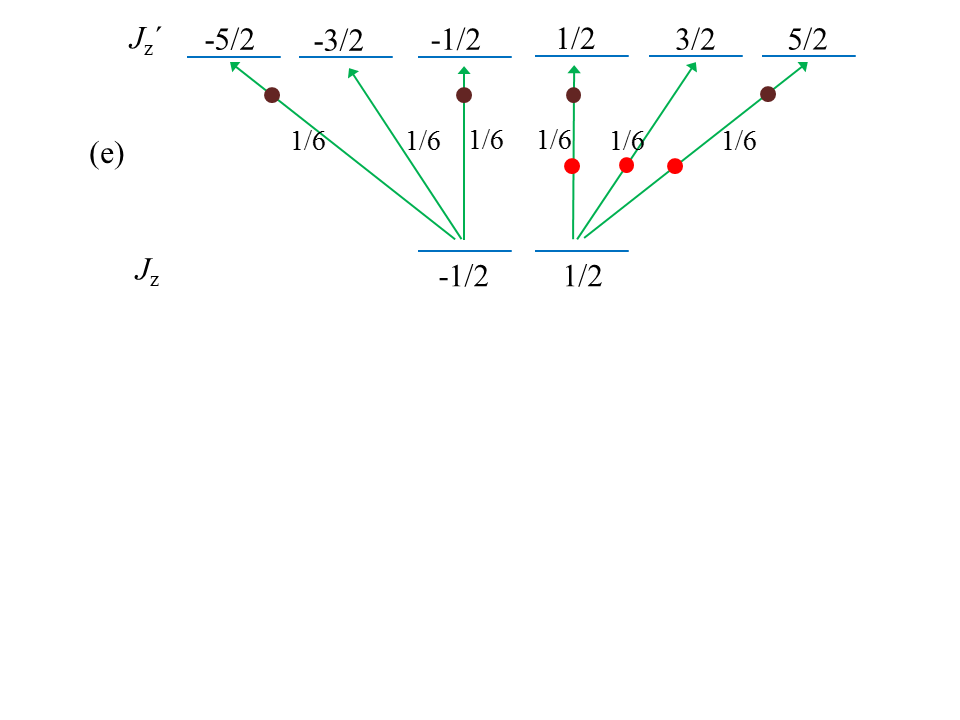}
\par\end{centering}
\centering{}\caption{\label{fig:magnetic and el.quadrupole transition diagram} Schematic
energy diagrams comprising a lower level havingangular momentum $J$
and an upper level $J'$. (a-e) show different cases $(J,\,J')$.
The green arrows denote transitions between specific magnetic sub-levels
$J_{z},\,J_{z}'$ to be measured. Each assigned coefficient $\alpha$
multiplies the corresponding transition frequency so that in the weighted
sum $f_{{\rm t}}^{1}$ the external-field perturbations average to
zero. (a -d): E1 or E2 transitions; (e): E2 transitions. The case
$J=2,\,J'=2$ is not shown; it can easily be obtained by generalization
of (a). The coefficients for the cases where $J>J'$ are obtained
from the above cases by symmetry.}
\end{figure}

(2) For each of these transitions, one measures the $J_{z}\rightarrow J_{z}'$
components as shown in Fig.~\ref{fig:magnetic and el.quadrupole transition diagram},
multiplies the components' frequencies with the weights indicated,
and further multiplies each with the factor $\alpha=(2J'+1)/{\cal N}'$.
The resulting weights are the $\tilde{\alpha}$ of eq.~(\ref{eq:f_t,tot}). 

$f_{{\rm t}}^{{\rm pert-free}}$ is free from LZ shift and EQ shift
by construction, already at the level of each individual transition
$FSJ\rightarrow F'S'J'$, because according to Fig.~\ref{fig:magnetic and el.quadrupole transition diagram},
(i) equal weight are given to the $J_{z}\rightarrow J_{z}'$ and $-J_{z}\rightarrow-J_{z}'$
transitions, (ii) all $J_{z}$ and $J_{z}'$ sub-states contribute
with equal weight. Furthermore, $f_{{\rm t}}^{{\rm pert-free}}$ is
free of spin structure because of the definition of weights as given
in (2). Finally, with this procedure, the total QZ shift also cancels:
because each $J_{z}'$ state enters $f_{{\rm t}}^{{\rm pert-free}}$
with the same (total) weight $1/(2|J_{z}'|+1){\cal N}'$ and similarly
for each $J_{z}$ state, eq.~(\ref{eq:Nulling of quadratic ZS})
applies. As an example, consider the $J_{z}'=-1$ states in Fig.~\ref{fig:magnetic and el.quadrupole transition diagram}~(a,b,c):
in each panel, their total weight is always 1/3. Thus,

\begin{equation}
h\,f_{{\rm t}}^{{\rm pert-free}}(p\rightarrow p')=E_{p'}^{0}-E_{p}^{0}\,.\label{eq:f_t,tot result}
\end{equation}
\textit{Thus, the traceless frequency defined in this way is equal
to the unperturbed (spin-averaged) frequency} $f^{0}$. This scheme
is generally applicable, but variations can be more efficient in terms
of minimizing the number of transitions to be measured, and will be
discussed below.

\subsection{Special cases}

For systems with integer $J$, there exist the $J_{z}=0\rightarrow J_{z}'=0$
Zeeman components. Measuring only these for the traceless frequency
leads to zero total LZ shift. 

Similarly, for systems with half-integer spin, the two Zeeman components
$J_{z}=1/2\rightarrow J_{z}'=-1/2$ and $J_{z}=1/2\rightarrow J_{z}'=-1/2$
exist. In the mean frequency of these two components the LZ effect
vanishes, for each hyperfine transition $FSJ\rightarrow F'S'J'$.

Such traceless frequencies will have a well-defined, nonzero EQ shift
and QZ shift. We shall take up this again in Sec.~\ref{subsec:Partial-cancellation}.\clearpage{}

\clearpage{}\pagebreak{}

\section{Application to the Molecular Hydrogen Ions\label{sec:Application-to-the}}

\subsection{Motivation}

The scope of the following discussion is to present in a detailed
manner the application of the traceless frequency to the elimination
of the spin-dependent energies in ${\rm H}_{2}^{+}$ and ${\rm HD}^{+}$.

Consider a single spin component $m\rightarrow m'$ within the spin
structure of a transition between the rovibrational levels $p:\,(v,\,N)\rightarrow p':\,(v',N')$,
($v$: vibrational quantum number, $N$: rotational quantum number).
The lower and upper spin state are enumerated by quantum numbers denoted
collectively as $m$ and $m'$. The frequency of an individual transition
is computed as a sum of two contributions: 
\begin{equation}
f(vNm\rightarrow v'N'm')=f^{{\rm spin-avg}}(v,N\rightarrow v',N')+f^{{\rm spin}}(v,N,m\rightarrow v',N',m')\,.
\end{equation}
The spin-averaged frequency $f^{{\rm spin-avg}}$ (corresponding to
$f^{0}$ above) depends only on $v$, $N$, $v'$, $N'$. Currently,
it can be computed with a fractional inaccuracy due to theory at the
$1\times10^{-11}$ level, for both vibrational \cite{Korobov2017a}
and rotational transitions \cite{Alighanbari2018}. The second contribution
is the spin-structure (hyperfine) shift $h\,f^{{\rm spin}}=E_{vN,m}^{{\rm spin}}-E_{v'N',m'}^{{\rm spin}}$
due to spin interactions. The effective spin hamiltonian $H^{{\rm spin}}$
has been derived within the Breit-Pauli approximation \cite{Bakalov2006,Korobov2006}.
It contains a set $\{{\cal E}_{q}\}$ of $(v,\,N)$ - dependent coefficients
, The number of coefficients in the set is up to 9 for ${\rm HD^{+}}$
and up to 5 for ${\rm H}_{2}^{+}$. The inaccuracy of the ab-initio
calculation of the set of coefficients is the dominant source of the
theoretical inaccuracy of $f_{{\rm spin}}$, at present \cite{Korobov2016}.

In general, the spin energies of a state are obtained numerically
by diagonalizing the spin hamiltonian. For ${\rm HD^{+}}$, only a
few particular states have energies expressible in explicit form.
This is the case for the stretched states (states of maximum total
angular momentum, $J_{{\rm max}}=N+2$, and maximum total angular
momentum projection, $|J_{z}|=J_{{\rm max}}$), for which (see eq.~(6)
in \cite{Bakalov2011includingerratum}),
\begin{align}
E_{vN}^{{\rm spin}}(F=1,\,S=2,\,J=N+2,\,J_{z}=\pm J)/h={\cal E}_{4}/4+{\cal E}_{5}/2\label{eq:E_spin of stretched states}\\
+({\cal E}_{1}+{\cal E}_{2}+2{\cal E}_{3}+{\cal E}_{6}+2{\cal E}_{7}+2{\cal E}_{8}+{\cal E}_{9})\,N/2\nonumber \\
-(2{\cal E}_{6}+4{\cal E}_{7}+4{\cal E}_{8}+2{\cal E}_{9})\,N^{2}/2\,,\nonumber 
\end{align}
where ${\cal E}_{q}={\cal E}_{q}(v,\,N)$ are the coefficients of
the effective spin Hamiltonian \cite{Bakalov2006,Bakalov2011includingerratum}.
This expression is helpful in showing how any theoretical inaccuracy
of the ${\cal E}_{q}$ will affect the inaccuracy of the overall transition
frequency $f$. 

Since the present considerations are independent of the particular
vibrational levels, we shall often omit the mention of $v,\,v'$ in
the state designations when they are not essential. 

\subsection{The molecular ion ${\rm HD}^{+}$\label{subsec:The-molecular-ion HDplus}}

Each state of ${\rm HD^{+}}$ is uniquely defined by the quantum numbers
$p$:~$(v,\,N)$ and $m$:~$(FSJJ_{z})$. As introduced in Sec.~\ref{sec:Cancellation-of-Spin Structure Contributions},
$F$ is (approximate or exact) spin of the electron - proton pair,
$S$ is the (approximate or exact) total spin of the three particles,
and $J$ is the (exact) total angular momentum quantum number of the
molecule. 

Before proceeding we note that the tracelessness property of the Breit-Pauli
interaction hamiltonian and the sum rule for the QZ shifts are relationships
that can be verified on computed energy shifts in order to check for
correctness of the computation \footnote{For example, an analytical calculation of the diagonal matrix elements
of the Breit-Pauli Hamiltonian can be performed in the basis of the
states $(I_{p},\,I_{p,z})$, $(I_{d},\,I_{d,z})$, $(s_{e},\,s_{e,z})$,
$(N,\,N_{z})$ with $I_{p}=1/2$, $I_{d}=1$, $s_{e}=1/2$. Taking
the sum over these elements must yield zero. One can also verify tracelessness
for many rovibrational levels computed in previous work, by performing
the weighted average, eq.~(\ref{eq:tracelessness for energies}),
over the numerical values of the spin energies given in Table~III
of ref.~\cite{Bakalov2006}. The sum rule for the QZ shifts can also
easily be verified numerically as follows. By summing over all entries
in the middle lines of Table~2 in ref.~\cite{Bakalov2011includingerratum}
(the values $q^{vLn}$) one verifies the case $J_{z}=0$. The cases
$J_{z}\ne0$ can be verified by appropriate sums over the middle and
lower lines together, omitting those entries belonging to states having
$J<|J_{z}|$. The sum rule for the QZ shifts can also easily be verified
numerically as follows. By summing over all entries in the middle
lines of Table~2 in ref.~\cite{Bakalov2011includingerratum} (the
values $q^{vLn}$) one verifies the case $J_{z}=0$. The cases $J_{z}\ne0$
can be verified by appropriate sums over the middle and lower lines
together, omitting those entries belonging to states having $J<|J_{z}|$.}. 

Let us initially ignore the external-field shifts. A general argument
allows us to find the weights $\alpha$ for the traceless frequency.
For simplicity, we assume zero magnetic field and electric field gradient,
so that the states are degenerate in $J_{z}$, and we omit this quantum
number in the following. The appropriate type of transitions are electric-dipole
(E1), which allow $J\rightarrow J'=J-1,\,J,\,J+1$ and $\Delta N=\pm1$.
Strong E1 transitions are those for which the conditions $\Delta F=0$
and $\Delta S=0$ are fulfilled. 

\begin{figure}[b]
\includegraphics[width=10cm]{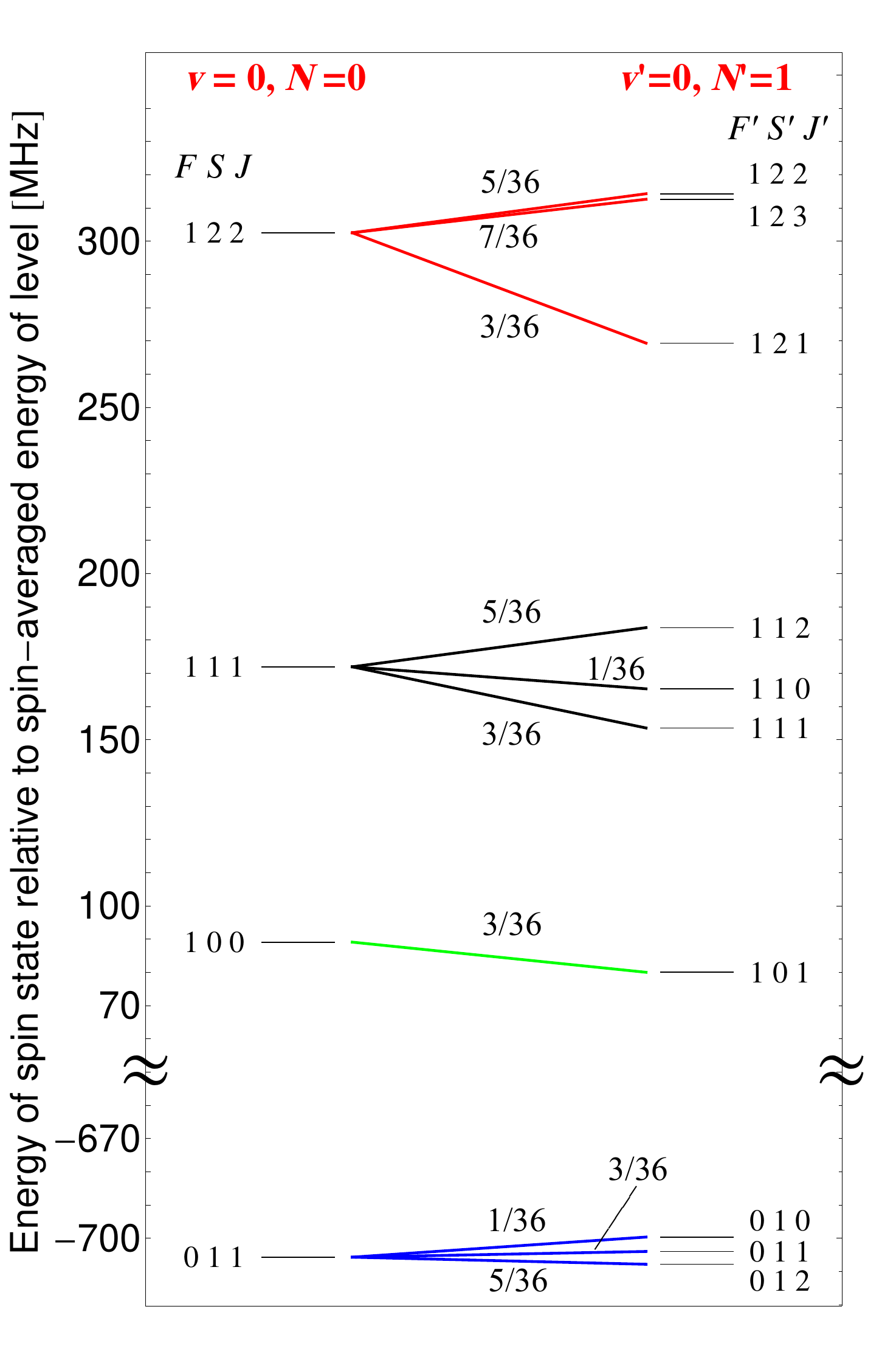}\caption{\label{fig:Spinenergies of rotational transition}The energies $E_{vN,m}^{{\rm spin}}$,
$E_{v'N',m'}^{{\rm spin}}$ of the spin states of ${\rm HD}^{+}$
relevant for the fundamental rotational transition $(v=0,\,N=0)\rightarrow(v'=0,\,N'=1)$.
The colored lines indicate the transitions to be measured in order
to compute the traceless transition frequency. The numbers above the
colored lines are the weights $\alpha$. For both levels, the zero
of the vertical scale corresponds to the respective spin-averaged
energy $E_{vN}^{{\rm spin-avg}}$. Adapted from ref.~\cite{Shen2012}.}
\end{figure}

For concreteness, we discuss the case of transitions from a lower
level $N=0$ to an upper level $N'=1$. There is no restriction on
$v$ and $v'$. Two transitions of this type have already been measured
with $10^{-9}$- level fractional inaccuracy \cite{Bressel2012,Alighanbari2018}.
One relevant case is the rotational transition between the two lowest-energy
rovibrational levels, $(v=0,\,N=0)\rightarrow(v'=0,\,N'=1)$, with
($f^{{\rm spin-avg}}\simeq1.3\,{\rm THz}$). The spin structure of
this transition has been discussed previously \cite{Bakalov2011includingerratum,Shen2012,Alighanbari2018}
and Fig.~\ref{fig:Spinenergies of rotational transition} reports
the detailed energy diagram with the actual spin energies $E^{{\rm spin}}$.

Each spin state $|N=0,\,F\,S\,J\,\rangle$ of the lower level having
$S\ne0$, can be excited to three spin states (forming a ``spin group'')
of the upper level, namely $|N'=1,\,F\,S\,J-1\rangle$, $|1,\,F\,S\,J\rangle$,
$|1,\,F\,S\,J+1\rangle$, by strong (allowed) transitions. To these
we assign weights proportional to the degeneracies of the respective
upper states, $\alpha(J;J')=(2J'+1)/{\cal N}'$. In the weighted sum
over all of these transitions in the traceless frequency, each state
of the upper level occurs once and therefore the total contribution
of the spin-averaged energies of the upper states yields $E^{{\rm spin-avg}}(v',N')/h$.
The total contribution of their spin energies however average to zero
because of the tracelessness of the spin hamiltonian. The state $|N=0,\,F=1,\,S=0,\,J=0\rangle$
is special: it has only a single strong transition, to the state $|N'=1,\,F'=1,\,S'=0,\,J'=1\rangle$
(green line in the figure). Nevertheless, the above weight assignment
is suitable. 

As a consequence of this weight assignment, the total weight of each
lower spin state in the sum is $\alpha(J;J-1)+\alpha(J;J)+\alpha(J;J+1)=3(2J+1)/{\cal N}'$,
since three transitions start from each state. This total weight naturally
turns out to be the degeneracy of the lower state, up to a constant
factor. Here again, ${\cal N}'=2\times2\times3\times(2N'+1)=36$ is
the number of states in the upper level. The special state $|0,\,F=1\,S=0\,J=0\rangle$
is again taken into account correctly, even if only a single transition
starts from it. The total contribution of the 4 lower states yields
\begin{align}
\sum_{FSJ\epsilon\{011,100,111,122\}}[3(2J+1)/{\cal N}']E_{N}^{{\rm spin-avg}}/h & =(3{\cal N}/{\cal N}')E_{N}^{{\rm spin-avg}}/h\\
 & =E_{N}^{{\rm spin-avg}}/h\,.
\end{align}
 Here, the number of states in the lower level is ${\cal N}=12$. 

Explicitly, the traceless frequency is (with the arbitrary vibrational
quantum numbers $v,\,v'$ reintroduced)

\begin{align}
f_{{\rm t}}^{{\rm spin}} & =\sum_{FSJ\rightarrow FSJ'}[(2J'+1)/{\cal N}']\,f(v,\,0,\,F\,S\,J\rightarrow v',\,1\,,F\,S\,J')\label{eq:traceless frequency for the rotational transition}\\
 & =f^{{\rm spin-avg}}(v,\,0\rightarrow v',\,1)\,.\nonumber 
\end{align}
The summation is limited to the 10 strong transitions (the colored
lines in Fig.~\ref{fig:Spinenergies of rotational transition}).
The traceless frequency eliminates the contributions from 11 spin
structure coefficients (9 of the upper level, 2 of the lower level),
for any choice of vibrational levels $v,\,v'$. For given $v,\,N,\,v',\,N'$,
the transition frequencies $f(v',N',m',\,v,N,m)$ lie in a range of
several 10~MHz, and thus only a single radiation source is sufficient
for their measurement. 

Transitions between levels whose rotational quantum numbers $N,\,N'$
are both nonzero require a generalization of eq.~(\ref{eq:traceless frequency for the rotational transition}).
The situation is now richer in the sense that in the lower level there
will typically be more than just one spin state $J$ for a given quantum
number pair $(F,\,S)$ in the lower level (see the right-hand side
of Fig.~\ref{fig:Spinenergies of rotational transition} for the
case $N=1$). This situation can easily be treated by setting up a
set of equations for the unknown weights $\alpha$, requiring that
the sum of weights of the strong spin transitions connecting to any
particular state $m$ or $m'$ be equal to the normalized Zeeman degeneracy
of that state, $(2J(m)+1)/{\cal N}$ or $(2J(m')+1)/{\cal N}$', respectively.
The set contains one equation for every state in the lower and in
the upper level. The solution of the set of equations shows that for
a $N=1\rightarrow N'=2$ transition, 18 spin transitions (5, 1, 5,
7 for the four spin groups, respectively) must be measured, and for
$N\ge2\rightarrow N'=N+1$ the number increases to 20 spin transitions
(5, 1, 5, 9 for the four spin groups, respectively). It is found that
not all strong transitions necessarily must contribute to $f_{{\rm t}}$.
The overall result is that in the traceless frequency, the influence
of 18 coefficients of the Breit-Pauli interaction is cancelled.

\subsection{The molecular ion ${\rm H}_{2}^{+}$\label{subsec:The-molecular-ion H2plus}}

${\rm H}_{2}^{+}$ exhibits some important differences compared to
${\rm HD}^{+}$ because it is homonuclear. States are denoted by $|v,\,N,\,I,\,F,\,J\rangle$,
where $I$ is the (exact) total nuclear spin quantum number, and $F$
is the (approximate or exact) total particle spin angular momentum
quantum number. 

Rovibrational levels with even $N=0,\,2,\,4,...$ are para levels
with zero total nuclear spin $I=0$. The total particle spin angular
momentum is $F=1/2$. The spin hamiltonian reduces to the spin-rotation
interaction, $H^{{\rm spin}}=c_{e}({\bf s_{e}\cdot N})$. Rovibrational
levels $(v,\,N)$ are therefore split into two if $N\ge2$. The energies
of the $J=N-1/2$ and $J=N+1/2$ levels are $-(N+1)c_{e}/2$ (if $N\ge2)$
and $Nc_{e}/2$, respectively.  

In the case of odd $N=1,\,3,...$ the molecule is in an ortho ($I=1$)
state and the total particle spin angular momentum is $F=1/2$ or
$F=3/2$. The number of spin levels is higher, 5 for $L=1$, and 6
for $L=3,\,5,...$. Fig.~\ref{fig:H2+ energy level scheme} shows
the spin structure of the lowest rovibrational levels.

As a homonuclear molecule, ${\rm H}_{2}^{+}$ cannot be interrogated
by one-photon electric-dipole (E1) transitions. Therefore, compared
to the ${\rm HD^{+}}$- case, an adapted discussion is required. Accessible
transitions are two-photon transitions and electric-quadrupole (E2)
transitions, already discussed in detail \cite{Karr2008a,Korobov2018a}.
We consider here only E2 transitions, because they show greater potential
than two-photon transitions. Fig.~\ref{fig:H2+ energy level scheme}
shows E2 transitions relevant for the following discussion. The crucial
issue are the selection rules for E2 transitions. The total molecular
angular momentum can change by $\Delta J=\text{0,\,\ensuremath{\pm}1,\,\ensuremath{\pm}2}$.
In particular, the fact that $\Delta J=0$ are allowed transitions
is important in the context of the present discussion. Such transitions
also have an unsuppressed strength, if $\Delta F=0$ \cite{Korobov2018a},
therefore they are experimentally accessible. An additional selection
rule is that $J=1/2\rightarrow J'=1/2$ is forbidden. Such transitions
could hypothetically only occur in the case $N=1\rightarrow N'=1$,
see also Fig.~\ref{fig:H2+ energy level scheme}. This case must
be treated carefully. We now discuss the cases of transitions between
para states and between ortho states separately.

\subsubsection{Para-${\rm H}_{2}^{+}$}

\begin{figure}[b]
\includegraphics[width=15cm]{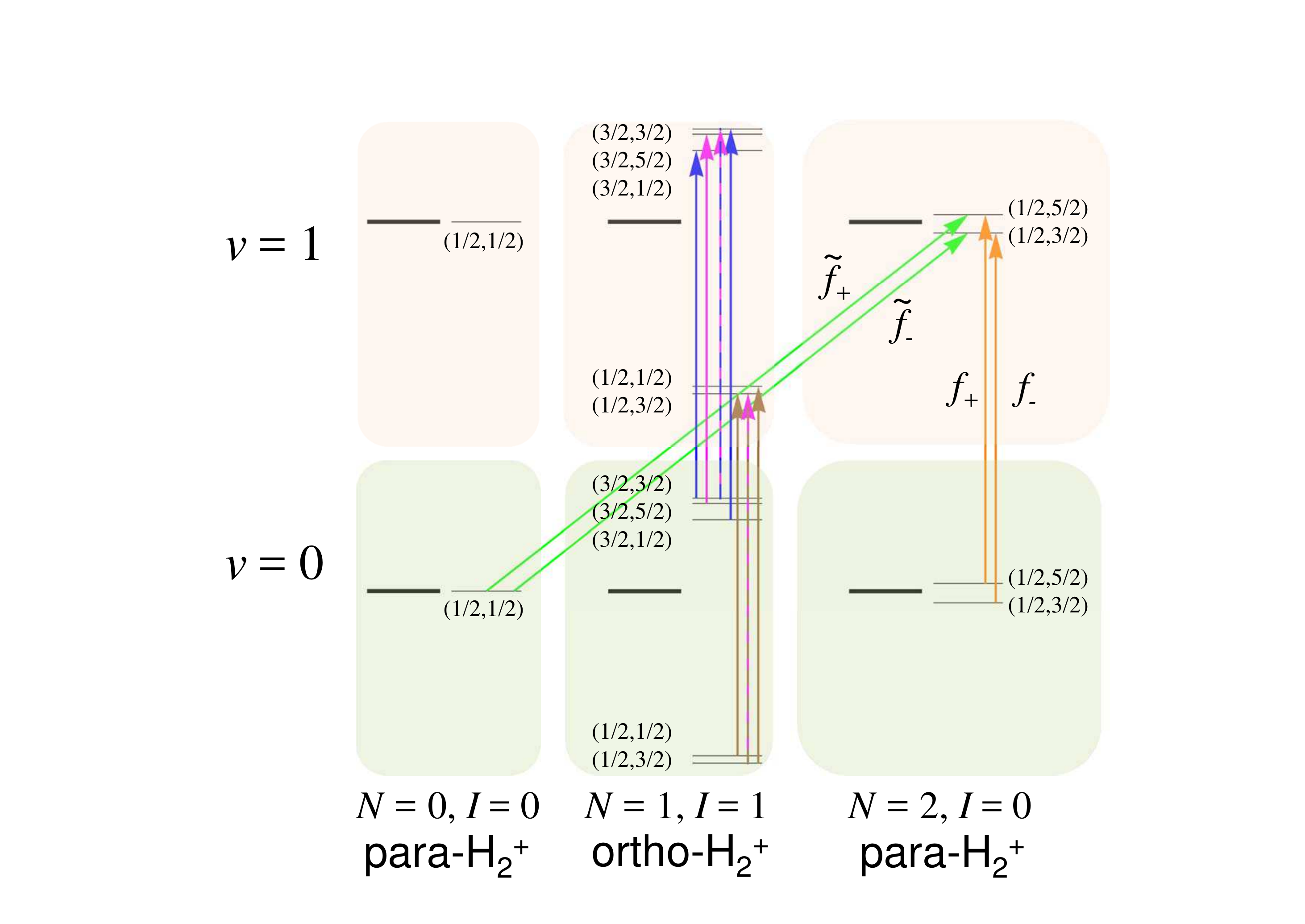}\caption{\label{fig:H2+ energy level scheme}Schematic of the spin structure
of the lowest rovibrational levels of ${\rm H}_{2}^{+}$. The number
pairs in parentheses are $(F,\,J)$. The thick lines represent the
spin-averaged energies of the levels. Energy splittings are not to
scale. The colored arrows denote transition frequencies to be measured.
The weights $\alpha$ are not indicated but are discussed in the text.}
\end{figure}

The spin states of para levels ($I=0$) are simple, pure angular momentum
states,
\begin{align*}
N & =0:\qquad|N,\,F=1/2,\,J=1/2\rangle,\\
N=2,\,4,...: & \qquad|N,\,F=1/2,\,J=N-1/2\rangle,\,|N,\,F=1/2,\,J=N+1/2\rangle\,.
\end{align*}
For transitions with $N\rightarrow N'=N$, the traceless frequency
is the weighted sum of \textit{two} frequencies ($f_{-},\,f_{+}$)
corresponding to $\Delta F=0$, $\Delta J=0$ - transitions, which
are shown as orange arrows in Fig.~\ref{fig:H2+ energy level scheme}:
\begin{align}
f_{{\rm t}}^{{\rm spin}} & =[(2J_{-}+1)\,f_{-}+(2J_{+}+1)\,f_{+}]/{\cal N}'\label{eq:f_c for para, general case}\\
 & =f^{{\rm spin-avg}}\,.\nonumber 
\end{align}
Here, $J_{\pm}=N\pm1/2$. The weights of $f_{-}$ and $f_{+}$ are
$N/(2N+1)$ and $(N+1)/(2N+1)$, respectively. The traceless frequency
eliminates the effect of the two relevant spin structure coefficients,
$c_{e}(v,\,N)$ for the lower level and $c_{e}(v',\,N'=N)$ for the
upper level.

If one wants to address levels with $N=0$ one must take into account
that transitions with $N=0\rightarrow N'=0$ are forbidden. In this
case, the traceless frequency is the weighted sum of \textit{two}
transitions with $\Delta N=+2$ (or $-2$), which start or end at
a \textit{common single} state. Now, the two transitions, denoted
by $\tilde{f}_{-},\,\tilde{f}_{+}$ have $\Delta J=1$ and $\Delta J=2$,
respectively. We consider the case $N=0,\,N'=2$, which is indicated
as green arrows in Fig.~\ref{fig:H2+ energy level scheme}. The spin
energy of the $N=0$ - level is zero, therefore

\begin{align}
f_{{\rm t,2}}^{{\rm spin}} & =[(2J_{-}'+1)\,\tilde{f}_{-}+(2J_{+}'+1)\,\tilde{f}_{+}]/{\cal N}'\label{eq: f_c for para, special case}\\
 & =f^{{\rm spin-avg}}\,.\nonumber 
\end{align}
The weights of $\tilde{f}_{-}$ and $\tilde{f}_{+}$ are 4/10 and
6/10, respectively. The traceless frequency eliminates the effect
of the single spin coefficient present in the problem, $c_{e}(v',N'=2)$
for the upper level. 

\subsubsection{Ortho-${\rm H}_{2}^{+}$\label{subsec:ft^spin for Ortho-H2+}}

For transitions between ortho levels ($I=1$) we shall limit ourselves
to the case $N\rightarrow N'=N$, which we consider the most experimentally
relevant at this time. Thus, the spin structure is the same in the
initial and final levels. The number of states is ${\cal N=N}'=6(2N+1)$.
The traceless frequency is the weighted sum over all transitions with
$\Delta F=0$, $\Delta J=0$,

\begin{align}
f_{{\rm t}}^{{\rm spin}} & =\sum_{J}[(2J+1)/{\cal N}]\,f(v,N,F,J\rightarrow v',N,F,J)\label{eq:f_c for ortho}\\
 & =\sum_{J}[(2J+1)/{\cal N}](f^{{\rm spin-avg}}+E_{v'NFJ}^{{\rm spin}}-E_{vNFJ}^{{\rm spin}})\nonumber \\
 & =[6(2N+1)/{\cal N}]\,f^{{\rm spin-avg}}+\sum_{J}[(2J+1)/{\cal N}]\,E_{v'NFJ}^{{\rm spin}}-\sum_{J}[(2J+1)/{\cal N}]\,E_{vNFJ}^{{\rm spin}}\nonumber \\
 & =f^{{\rm spin-avg}}(v,N\rightarrow v',N)\,.\nonumber 
\end{align}
The sum includes 5 transitions if $N=1$ and 6 transitions if $N=3,\,5,\ldots$
The traceless frequency eliminates the effect of 10 spin coefficients,
5 for the lower level, and 5 for the upper level.

As mentioned, the case $N=1\rightarrow N'=1$ is special because both
the initial and the final rovibrational levels include two spin states
having total angular momentum $J=1/2$ and the $\Delta J=0$ - transitions
between these are forbidden. Thus, two frequencies in the first sum
in eq.~(\ref{eq:f_c for ortho}), $f(N,\,F=1/2,\,J=1/2\rightarrow N,\,F,\,J'=J)$
and $f(N,\,F=3/2,\,J=1/2\rightarrow N,\,F,\,J'=J)$, cannot be experimentally
accessed.This problem is solved by determining these via a combination
of allowed transitions. Several such combinations are possible; one
of them is:
\begin{align}
f(N,\,F=1/2,\,J=1/2\rightarrow N,\,F,\,J'=J)= & f(N,\,1/2,\,1/2\rightarrow N,\,1/2,\,3/2)\label{eq:Special case of Jz=00003D1/2 to Jz=00003D-1/2}\\
 & -f(N,\,1/2,\,3/2\rightarrow N,\,1/2,\,3/2)\nonumber \\
 & +f(N,\,1/2,\,3/2\rightarrow N,\,1/2,\,1/2)\,.\nonumber 
\end{align}
These three transitions are the two brown arrows and the brown-magenta
dashed arrow in the figure. Similarly, $f(N,\,F=3/2,\,J=1/2\rightarrow N,\,F,\,J'=J)$
can be determined via the two blue arrows and the blue-magenta dashed
arrow in the figure. The remaining three frequencies $f(F=1/2,\,J=3/2\rightarrow F,\,J)$,
$f(F=3/2,~J=3/2\rightarrow F,\,J)$ and $f(F=3/2,~J=5/2\rightarrow F,\,J)$
are accessible and are shown as the magenta, the magenta-brown dashed,
and the magenta-blue dashed arrows in Fig.~\ref{fig:H2+ energy level scheme}.
Because there is a partial overlap in the transitions to be measured,
only 2 additional frequencies are required in order to overcome the
$J=1/2\not\rightarrow J'=1/2$ selection rule, yielding a total of
7 transitions.

All considered transition frequencies have $\Delta F=0$, thus they
lie within a range of approximately 200~MHz for a given rovibrational
level, requiring only a single laser for their measurement.  

\section{Canceling the Systematic Shifts\label{subsec:Canceling-the-Systematic Shifts}}

\subsection{Partial cancellation\label{subsec:Partial-cancellation}}

Line~1 in Tab.~\ref{tab:Number of required components-1} summarizes
the discussion of Sec.~\ref{sec:Application-to-the}, the case of
traceless frequency where the Zeeman components and thus the systematic
shifts are not considered. Line 2 is the case where the LZ shift is
canceled, but all other shifts are not canceled, except in particular
cases. We denote this traceless frequency by $f_{{\rm t}}^{{\rm spin,LZ}}$.
It will be discussed next.
\begin{center}
\begin{sidewaystable}
\begin{tabular*}{1\columnwidth}{@{\extracolsep{\fill}}|>{\centering}m{0.2\columnwidth}||>{\centering}m{0.2\columnwidth}|>{\centering}p{0.2\columnwidth}|>{\centering}p{0.3\columnwidth}|}
\hline 
\multirow{2}{0.2\columnwidth}{cancellation of} & \multicolumn{2}{c|}{${\rm H}_{2}^{+}$~(E2)} & ${\rm HD}^{+}$~(E1)\tabularnewline
\cline{2-4} 
 & $N=0\rightarrow N'=2$ & $N=2\rightarrow N'=2$ & $N=0\rightarrow N'=1$\tabularnewline
\hline 
\hline 
(1) SS $(f_{{\rm t}}^{{\rm spin}})$ & $1\left(\frac{1}{2}\rightarrow\frac{3}{2}\right)+$ $1\left(\frac{1}{2}\rightarrow\frac{5}{2}\right)=2$ & $1\left(\frac{3}{2}\rightarrow\frac{3}{2}\right)+$ $1\left(\frac{5}{2}\rightarrow\frac{5}{2}\right)=2$ & 10\tabularnewline
\hline 
(2) SS + LZS $(f_{{\rm t}}^{{\rm spin,LZ}})$ & $2\left(\frac{1}{2}\rightarrow\frac{3}{2}\right)+$$2\left(\frac{1}{2}\rightarrow\frac{5}{2}\right)=4${*} & $2\left(\frac{3}{2}\rightarrow\frac{3}{2}\right)+$ $2\left(\frac{5}{2}\rightarrow\frac{5}{2}\right)=4${*} & 10\tabularnewline
\hline 
\multirow{1}{0.2\columnwidth}{(3) SS + LZS + QZS + EQS $(f_{{\rm t}}^{{\rm pert-free}})$} & \multirow{1}{0.2\columnwidth}{$4\left(\frac{1}{2}\rightarrow\frac{3}{2}\right)+$ $6\left(\frac{1}{2}\rightarrow\frac{5}{2}\right)=10${*}{*}} & \multirow{1}{0.2\columnwidth}{$4\left(\frac{3}{2}\rightarrow\frac{3}{2}\right)+$ $6\left(\frac{5}{2}\rightarrow\frac{5}{2}\right)=10${*}{*}} & $\begin{array}{c}
9+5+15\,(2\rightarrow1,2,3)+\\
3+3+9\,(1\rightarrow0,1,2)+\\
+3\,(0\rightarrow1)+\\
3+3+9\,(1\rightarrow0,1,2)=\\
62\,{\rm **}
\end{array}$\tabularnewline
\hline 
\end{tabular*}\caption{\label{tab:Number of required components-1}Overview of the number
$n$ of Zeeman components to be measured in order to achieve traceless
frequencies with different cancellation of systematic shifts. The
expressions in parentheses denote the contributing transitions $J\rightarrow J'$.
The asterisk ({*}) indicates that there exist particular choices that
in addition have zero QZS. The double asterisks ({*}{*}) indicate
cases where more efficient strategies can achieve lower number of
transitions to be measured, see Sec.~\ref{subsec:Largely-complete-cancellation}.
SS: spin structure; LZS: linear Zeeman shift; QZS: quadratic Zeeman
shift; EQS: electric quadrupole shift.}
\end{sidewaystable}
\par\end{center}

\subsubsection{${\rm H}_{2}^{+}$}

Cancellation of LZ shift in the traceless frequency is implemented,
for ${\rm H}_{2}^{+}$, by measuring, for each $J\rightarrow J'$
transition, the pair $J_{z}\rightarrow J_{z}'$, $-J_{z}\rightarrow-J_{z}'$.
The values of $J_{z},\,J_{z}'$ can be chosen freely. Thus, there
is a substantial number of options for implementation. This requires
doubling the number of spin components to be measured, compared to
the case of line 1. For transitions between the para states of ${\rm H}_{2}^{+}$
shown in Fig.~\ref{fig:H2+ energy level scheme} (green or orange
arrows), the number is 4, still small. For such transitions, the resulting
EQ, Stark, and QZ shifts are reported in Tab.~\ref{tab:H2+: Examples-of-traceless frequencies with only one pair of Zeeman components per hyperfine transition-1}.
The table shows all 6 possible options for the $(0,0)\rightarrow(1,2)$
transition, and a subset of all options for the $(0,2)\rightarrow(1,2)$
transition. 

The EQ shift in para states of ${\rm H}_{2}^{+}$ was computed using
\[
E^{{\rm EQ}}(v,\,N,\,F,\,S,\,J,\,J_{z})=\frac{3}{2}\sqrt{\frac{3}{2}}E_{14}Q_{zz}(v,\,N)\left(-\frac{1}{2}\frac{D'(J(J+1)-3J_{z}^{2})}{3J(J+1)(2J-1)(2J+3)}\right)V_{zz}\,,
\]
where $D'=3D(D-1)-4J(J+1)N(N+1)$, $D=J(J+1)+N(N+1)-3/4$. This results
from eq.~(40) in \cite{Bakalov2014includingcorrigendum} and Eqs.~(10,~25)
in \cite{Schiller2014a}. 

In the table, we recognize two cases in which the QZ shift is zero
(underlined). Clearly, these are the transitions of choice. For an
electric field gradient assumed to be $V_{zz}\simeq0.07\,{\rm GV/m^{2}}$
, the shift of these two transitions is 5~Hz $(\simeq1\times10^{-13})$
and -0.6~Hz $(\simeq1\times10^{-14})$, respectively.

For the first transition, $(0,0)\rightarrow(1,2)$, the LZ shifts
of the contributing Zeeman transitions, weighted according to eq.~(\ref{eq: f_c for para, special case}),
are $\simeq-56$~kHz/G and $\simeq252$~kHz/G, respectively. This
means that in a magnetic field $B=0.1$~G, its stability must be
better than 1~part in 5000, if the residual LZ after averaging over
the Zeeman components is to be smaller than the EQ shift.

For the second transition, $(0,2)\rightarrow(1,2)$, the weighted
LZ shifts are approximately $10^{5}$ times smaller, $\simeq0.9$~Hz/G
and $\simeq2.5$~Hz/G, respectively. Thus, no high magnetic field
stability is needed. This is an attractive choice for high-precision
measurements.

The order of magnitude for the time-averaged electric field squared
in a macroscopic linear ion trap is $E^{2}\simeq1\,{\rm (kV/m)^{2}}$,
whereupon the DC Stark shift is approximately 10~mHz for all transitions
shown in the table, or $\le2\times10^{-16}$ fractionally.\clearpage{}
\begin{center}
\begin{table}[H]
\begin{centering}
\begin{tabular}{|c||c|c|c|c|c|}
\hline 
\multirow{2}{*}{Transition} & \multirow{2}{*}{$n$} & Zeeman & EQ shift/$V_{zz}$  & DC Stark shift/$E^{2}$  & QZ shift/$B^{2}$\tabularnewline
 &  & components & (Hz m$^{2}$/GV) & (Hz (m/kV)$^{2}$) & (kHz/${\rm G}^{2}$)\tabularnewline
\hline 
\hline 
$\text{E2:\,(0,0)\ensuremath{\to}(1,2)}$ & 4 & $\begin{array}{c}
\frac{1}{2}\rightarrow\frac{3}{2},\,\frac{1}{2}\rightarrow\frac{3}{2}\end{array}$ & 29.2 & $0.0033\cos^{2}\theta-0.011$ & 2.54\tabularnewline
\hline 
$\text{E2:\,(0,0)\ensuremath{\to}(1,2)}$ & 4 & $\begin{array}{c}
\frac{1}{2}\rightarrow\frac{3}{2},\,\frac{1}{2}\rightarrow\frac{1}{2}\end{array}$ & $-36.4$ & $-0.0041\cos^{2}\theta-0.0082$ & 6.35\tabularnewline
\hline 
$\text{E2:\,(0,0)\ensuremath{\to}(1,2)}$ & 4 & $\begin{array}{c}
\frac{1}{2}\rightarrow\frac{1}{2},\,\frac{1}{2}\rightarrow\frac{5}{2}\end{array}$ & 58.3 & $0.0066\cos^{2}\theta-0.012$ & -7.63\tabularnewline
\hline 
$\underline{\text{E2:\,(0,0)\ensuremath{\to}(1,2)}}$ & 4 & $\begin{array}{c}
\frac{1}{2}\rightarrow\frac{1}{2},\,\frac{1}{2}\rightarrow\frac{3}{2}\end{array}$ & $-72.9$ & $-0.0082\cos^{2}\theta-0.0068$ & 0\tabularnewline
\hline 
$\text{E2:\,(0,0)\ensuremath{\to}(1,2)}$ & 4 & $\begin{array}{c}
\frac{1}{2}\rightarrow\frac{1}{2},\,\frac{1}{2}\rightarrow\frac{1}{2}\end{array}$ & $-139$ & $-0.016\cos^{2}\theta-0.0043$ & 3.81\tabularnewline
\hline 
$\text{E2:\,(0,0)\ensuremath{\to}(1,2)}$ & 4 & $\begin{array}{c}
\frac{1}{2}\rightarrow\frac{3}{2},\,\frac{1}{2}\rightarrow\frac{5}{2}\end{array}$ & 160 & $0.018\cos^{2}\theta-0.016$ & -5.08\tabularnewline
\hline 
\hline 
$\text{E2:\,(0,2)\ensuremath{\to}(1,2)}$ & 4 & $\begin{array}{c}
\frac{3}{2}\rightarrow\frac{3}{2},\,\frac{3}{2}\rightarrow\frac{3}{2}\end{array}$ & $3.63$ & $0.00098\cos^{2}\theta-0.0095$ & $0.156$\tabularnewline
\hline 
$\text{E2:\,(0,2)\ensuremath{\to}(1,2)}$ & 4 & $\begin{array}{c}
\frac{3}{2}\rightarrow\frac{3}{2},\,\frac{1}{2}\rightarrow\frac{1}{2}\end{array}$ & $-4.53$ & $-0.0012\cos^{2}\theta-0.0088$ & $0.390$\tabularnewline
\hline 
$\text{E2:\,(0,2)\ensuremath{\to}(1,2)}$ & 4 & $\begin{array}{c}
\frac{1}{2}\rightarrow\frac{1}{2},\,\frac{5}{2}\rightarrow\frac{5}{2}\end{array}$ & $7.25$ & $0.0020\cos^{2}\theta-0.0098$ & $-0.469$\tabularnewline
\hline 
$\underline{\text{E2:\,(0,2)\ensuremath{\to}(1,2)}}$ & 4 & $\begin{array}{c}
\frac{1}{2}\rightarrow\frac{1}{2},\,\frac{3}{2}\rightarrow\frac{3}{2}\end{array}$ & $-9.07$ & $-0.0024\cos^{2}\theta-0.0084$ & $0$\tabularnewline
\hline 
\textbf{$\boldsymbol{\text{E2:\,(0,2)\ensuremath{\to}(1,2)}}$} & 4 & $\begin{array}{c}
\frac{1}{2}\rightarrow\frac{1}{2},\,\frac{1}{2}\rightarrow\frac{1}{2}\end{array}$ & $-17.2$ & $-0.0046\cos^{2}\theta-0.0076$ & $0.234$\tabularnewline
\hline 
\textbf{$\boldsymbol{\text{E2:\,(0,2)\ensuremath{\to}(1,2)}}$} & 4 & $\begin{array}{c}
\frac{3}{2}\rightarrow\frac{3}{2},\,\frac{5}{2}\rightarrow\frac{5}{2}\end{array}$ & $19.9$ & $0.0054\cos^{2}\theta-0.011$ & $-0.312$\tabularnewline
\hline 
$\text{E2:\,(0,2)\ensuremath{\to}(1,2)}$ & 4 & $\begin{array}{c}
\frac{1}{2}\rightarrow\frac{3}{2},\,\frac{5}{2}\rightarrow\frac{3}{2}\end{array}$ & $-21.9$ & $-0.0013\cos^{2}\theta-0.0087$ & $9.70$\tabularnewline
\hline 
$\text{E2:\,(0,2)\ensuremath{\to}(1,2)}$ & 4 & $\begin{array}{c}
\frac{1}{2}\rightarrow\frac{3}{2},\,\frac{3}{2}\rightarrow\frac{1}{2}\end{array}$ & $27.4$ & $0.0017\cos^{2}\theta-0.0097$ & $6.35$\tabularnewline
\hline 
$\text{E2:\,(0,2)\ensuremath{\to}(1,2)}$ & 4 & $\begin{array}{c}
\frac{3}{2}\rightarrow\frac{1}{2},\,\frac{3}{2}\rightarrow\frac{5}{2}\end{array}$ & $32.8$ & $0.0043\cos^{2}\theta-0.011$ & $-10.0$\tabularnewline
\hline 
$\text{E2:\,(0,2)\ensuremath{\to}(1,2)}$ & 4 & $\begin{array}{c}
\frac{3}{2}\rightarrow\frac{1}{2},\,\frac{1}{2}\rightarrow\frac{3}{2}\end{array}$ & $-41.0$ & $-0.0053\cos^{2}\theta-0.0074$ & $-5.96$\tabularnewline
\hline 
$\text{E2:\,(0,2)\ensuremath{\to}(1,2)}$ & 4 & $\begin{array}{c}
\frac{1}{2}\rightarrow\frac{1}{2},\,\frac{1}{2}\rightarrow\frac{3}{2}\end{array}$ & $48.4$ & $0.0027\cos^{2}\theta-0.010$ & $-3.58$\tabularnewline
\hline 
$\text{E2:\,(0,2)\ensuremath{\to}(1,2)}$ & 4 & $\begin{array}{c}
\frac{3}{2}\rightarrow\frac{1}{2},\,\frac{5}{2}\rightarrow\frac{1}{2}\end{array}$ & $-279$ & $-0.028\cos^{2}\theta+0.00025$ & $8.58$\tabularnewline
\hline 
$\text{E2:\,(0,2)\ensuremath{\to}(1,2)}$ & 4 & $\begin{array}{c}
\frac{1}{2}\rightarrow\frac{3}{2},\,\frac{1}{2}\rightarrow\frac{5}{2}\end{array}$ & $282$ & $0.029\cos^{2}\theta-0.019$ & $-8.66$\tabularnewline
\hline 
\end{tabular}
\par\end{centering}
\caption{\label{tab:H2+: Examples-of-traceless frequencies with only one pair of Zeeman components per hyperfine transition-1}${\rm H}_{2}^{+}$:
the systematic shifts contributions to traceless frequencies $f_{{\rm t}}^{{\rm spin,LZ}}$
which include only one pair of Zeeman components per hyperfine transition.
The spin structure contributions cancel and the linear Zeeman shift
is zero for all cases. For each rovibrational transition, the cases
are ordered according to an increasing absolute value of the EQ shift.
For the $(0,2)\rightarrow(1,2)$ transition, the first 11 and the
last 2 of the 36 possible cases are shown. $n$ is the number of transitions
that contribute to $f_{t}^{{\rm spin,LZ}}$. The third column reports
(omitting the signs) the values $\pm J_{z}\rightarrow\pm J_{z}'$
of the $J=1/2\rightarrow J'=3/2$, $J=1/2\rightarrow J'=5/2$ hyperfine
transitions (in the upper section) or of the $J=3/2\rightarrow J'=3/2$,
$J=5/2\rightarrow J'=5/2$ transitions (in the lower section). $\theta$
is the angle between the electric field vector and the magnetic field
vector (quantization axis). }
\end{table}
\par\end{center}

With the options given by Tab.~\ref{tab:H2+: Examples-of-traceless frequencies with only one pair of Zeeman components per hyperfine transition-1}
it is easy to consider the combination of two transitions, $f_{{\rm c}}=\beta_{1}f_{{\rm t},1}^{{\rm spin,LZ}}+\beta_{2}f_{{\rm t},2}^{{\rm spin,LZ}}$,
with the weights satisfying the condition $\beta_{i_{1}}+\beta_{i_{2}}+\ldots=1$
in order not to suppress the spin-averaged frequency. A total of 8
Zeeman components would need to be measured. Particularly attractive
is the combination of the last two transitions in the table. They
can be combined either to null the composite EQ shift, resulting in
a composite QZ shift of $0.0028\,{\rm kHz/G^{2}}$ or to null the
composite QZ shift, giving a composite EQ shift of $0.091\,{\rm Hz\,m^{2}/GV}$.
In practice, the nulling can only be as good as the stability of the
fields during measurement. It may then be more robust to combine the
11th and 12th transition (bold in the table), which have smaller shifts.
The individual Zeeman components contribute with very small LZ shifts
of $\simeq4$~Hz/G or less, so that magnetic field stability is not
an issue. If the combination is chosen to null the QZ shift, the
remaining dominant shift is the EQ shift. For the field strength value
assumed above, we may expect a fractional shift of approximately $2\times10^{-15}$.

Note that such combinations $f_{{\rm c}}$ contain only one transition
pair less than the general expression, i.e. 8 vs. 10 Zeeman components.
Even more efficient combinations will be presented below in Sec.~\ref{subsec:Largely-complete-cancellation}.

The smallness of the LZ shift and QZ shift of the individual Zeeman
components entering the 2\textsuperscript{nd} zero-QZ shift transition
(underlined in table) and the 11\textsuperscript{th} and 12\textsuperscript{th}
transitions was already emphasized earlier by us \cite{Schiller2014}.

\subsubsection{${\rm HD}^{+}$}

For cancellation of LZ shift in ${\rm HD}^{+}$, we can choose to
measure only the particular component $J_{z}=0\rightarrow J_{z}'=0$,
and the number of spin components to be measured remains the same
as in line 1. Table~\ref{tab:Examples-of-Jz=00003D0 traceless frequencies-1}
includes transitions with different $(N,\,N')$, and shows several
E1 transitions. The EQ shift was computed following Ref.~\cite{Bakalov2014includingcorrigendum}.
For the electric field gradient $V_{zz}\simeq0.07\,{\rm GV/m^{2}}$
and a typical magnetic field $B\simeq0.1\,{\rm G}$, the two E1 vibrational
transitions, $(0,0)\rightarrow(1,1)$, $(0,1)\rightarrow(1,2)$, have
a shift of approximately 150~Hz, or $2\times10^{-12}$, dominated
by QZ shifts. These absolute values are also the case for the two
pure rotational transitions. But since these have $\simeq45$ or 15~times
smaller frequency, the fractional shifts are $\simeq(1,\,0.3)\times10^{-10}$,
respectively. 

Table~\ref{tab:Examples-of-Jz=00003D0 traceless frequencies-1} also
includes electric quadrupole transitions (E2). It is unusual to consider
such transitions for heteronuclear diatomic molecules, but we see
from the table that for a given lower rovibrational level they reduce
significantly the number of Zeeman components to be measured. In addition,
for the vibrational transitions with $\Delta J=\pm2$ the QZ shifts
are reduced by a factor of approximately 10 compared to the E1 transitions.
For and the total shift is of order 10~Hz ($2\times10^{-13}$ fractionally),
still dominated by the QZ shift. For the $\Delta J=0$ transitions,
the QZ shift is even smaller, e.g. for the $(0,2)\rightarrow(1,2)$
transition, the total shift is $\simeq3$~Hz, or $5\times10^{-14}$.
This is an attractive transition for high-precision experiments. 
\begin{center}
\begin{table}[H]
\begin{centering}
\begin{tabular}{|c||c|c|c|r@{\extracolsep{0pt}.}l|}
\hline 
\multirow{2}{*}{Transition} & $n$ & EQ shift/$V_{zz}$  & DC Stark shift/$E^{2}$  & \multicolumn{2}{c|}{QZ shift/$B^{2}$}\tabularnewline
 &  & (Hz m$^{2}$/GV) & (Hz (m/kV)$^{2}$) & \multicolumn{2}{c|}{(kHz/${\rm G}^{2}$)}\tabularnewline
\hline 
\hline 
$\text{E1:\,(0,0)\ensuremath{\to}(0,1)}$ & 10 & $-37$ & $4.6+0.68\,\cos^{2}\theta$ & \multicolumn{2}{c|}{$-14$}\tabularnewline
\hline 
$\text{E1:\,(0,1)\ensuremath{\to}(0,2)}$ & 22 & $-48$ & $0.054-0.16\,\cos^{2}\theta$ & \multicolumn{2}{c|}{$15$}\tabularnewline
\hline 
$\text{E1:\,(0,0)\ensuremath{\to}(1,1)}$ & 10 & $-41$ & $4.6+0.79\,\cos^{2}\theta$ & \multicolumn{2}{c|}{$-15$}\tabularnewline
\hline 
$\text{E1:\,(0,1)\ensuremath{\to}(1,2)}$ & 22 & $-59$ & $0.016-0.076\,\cos^{2}\theta$ & \multicolumn{2}{c|}{$15$}\tabularnewline
\hline 
\hline 
$\text{E2:\,(0,0)\ensuremath{\to}(1,2)}$ & 12 & $-95$ & $4.7+0.60\,\cos^{2}\theta$ & \multicolumn{2}{c|}{$1.4$}\tabularnewline
\hline 
$\text{E2:\,(0,1)\ensuremath{\to}(1,1)}$ & 10 & $-4.6$ & $-0.047+0.12\,\cos^{2}\theta$ & \multicolumn{2}{c|}{$-0.75$}\tabularnewline
\hline 
$\text{E2:\,(0,2)\ensuremath{\to}(1,2)}$ & 12 & $-11$ & $-0.038+0.087\,\cos^{2}\theta$ & \multicolumn{2}{c|}{$0.27$}\tabularnewline
\hline 
$\text{E2:\,(0,2)\ensuremath{\to}(1,0)}$ & 12 & $85$ & $-5.5-0.52\,\cos^{2}\theta$ & \multicolumn{2}{c|}{$-1.0$}\tabularnewline
\hline 
\end{tabular}
\par\end{centering}
\caption{\label{tab:Examples-of-Jz=00003D0 traceless frequencies-1}The molecule
HD\protect\textsuperscript{+}: Examples of the systematic shifts
contributions to traceless frequencies $f_{{\rm t}}^{{\rm spin,LZ}}$
which include only the $J_{z}=0\rightarrow J_{z}'=0$ components for
each hyperfine transition, according to eq.~(\ref{eq:f_t,tot}).
The spin structure contributions cancel and the linear Zeeman shift
is zero for all cases.}
\end{table}
\par\end{center}

\subsection{Largely complete cancellation - general method}

According to the general formula eq.~(\ref{eq:f_t,tot}) we can combine
the spin structure cancellation (which did not consider the individual
Zeeman components $J_{z}\rightarrow J_{z}'$) with a cancellation
of the Zeeman and EQ shifts. This will typically require to increase
substantially the number of Zeeman components to be measured, eq.~\ref{eq:Nulling of quadratic ZS}.
Table~\ref{tab:Number of required components-1}, line 3, indicates
this number. It is obtained directly from the Fig.~\ref{fig:magnetic and el.quadrupole transition diagram}. 

An example will illustrate the cancellation of the quadratic Zeeman
shift. Consider ${\rm HD}^{+}$and the transition shown in Fig.~\ref{fig:Spinenergies of rotational transition}.
 The double sum over $m$, $m'$ in eq.~(\ref{eq:Nulling of quadratic ZS})
includes all $J_{z}$ and all $J_{z}'$. Consider the contributions
having a particular value of $J_{z}'$ or of $J_{z}$. With the help
of Fig.~\ref{fig:magnetic and el.quadrupole transition diagram}~(a,b,c)
we find the weights $\tilde{\alpha}$ in the traceless frequency $f_{{\rm t}}^{{\rm pert-free}}$,
to be assigned to the transitions connecting to the different $FSJ$,
$F'S'J'$ states. Table~\ref{tab:Weights--for the fundamental rotational transition in HD+}
shows two cases, $J_{z}'=-1$, and $J_{z}=0$. It is seen that the
QZ shift stemming from all transitions to each of these states is
indeed nulled.\clearpage{}
\begin{center}
\begin{table}[tb]
\begin{tabular}{|>{\centering}m{0.1\columnwidth}|c|>{\centering}p{0.05\columnwidth}|>{\centering}p{0.15\columnwidth}|>{\centering}p{0.15\columnwidth}|>{\centering}p{0.15\columnwidth}|}
\hline 
\multirow{3}{0.1\columnwidth}{lower level $FSJ$} & \multicolumn{5}{c|}{upper level}\tabularnewline
\cline{2-6} 
 & \multirow{2}{*}{$F'S'$} & \multicolumn{4}{c|}{$J'$}\tabularnewline
\cline{3-6} 
 &  & 0 & 1 & 2 & 3\tabularnewline
\hline 
122 & 12 & - & (b): $(3/36)\times(\frac{3}{15}+\frac{1}{15}+\frac{1}{15})$ & $(5/36)\times(\frac{1}{5})$ & (c): $(7/36)\times(\frac{1}{35}+\frac{1}{35}+\frac{3}{35})$\tabularnewline
\hline 
111 & 11 & - & (a):~$(3/36)\times(\frac{1}{3})$ & (b):~$(5/36)\times(\frac{1}{15}+\frac{2}{15})$ & -\tabularnewline
\hline 
100 & 10 & - & (a):~$(3/36)\times(\frac{1}{3})$ & - & -\tabularnewline
\hline 
011 & 01 & - & (a):~$(3/36)\times(\frac{1}{3})$ & (b):~$(5/36)\times(\frac{1}{15}+\frac{2}{15})$ & -\tabularnewline
\hline 
\end{tabular}\medskip{}

\medskip{}

\medskip{}

\begin{tabular}{|>{\centering}p{0.1\columnwidth}|c|>{\centering}p{0.15\columnwidth}|>{\centering}p{0.15\columnwidth}|>{\centering}p{0.15\columnwidth}|>{\centering}p{0.15\columnwidth}||c|}
\hline 
\multirow{3}{0.1\columnwidth}{lower level $FSJ$} & \multicolumn{5}{c||}{upper level} & \multirow{3}{*}{$\sum'_{m'}\tilde{\alpha}(p',m';p,m)$}\tabularnewline
\cline{2-6} 
 & \multirow{2}{*}{$F'S'$} & \multicolumn{4}{c||}{$J'$} & \tabularnewline
\cline{3-6} 
 &  & 0 & 1 & 2 & 3 & \tabularnewline
\hline 
122 & 12 & - & (b): $(3/36)\times(\frac{1}{15}+\frac{1}{15}+\frac{1}{15})$ & $(5/36)\times(\frac{1}{5})$ & (c): $(7/36)\times(\frac{3}{35}+\frac{1}{35}+\frac{3}{35})$ & $3/36$\tabularnewline
\hline 
111 & 11 & (a):~$(1/36)\times(\frac{1}{3})$ & (a):~$(3/36)\times(\frac{1}{3})$ & (b):~$(5/36)\times(\frac{2}{15}+\frac{1}{15}+\frac{2}{15})$ & - & $3/36$\tabularnewline
\hline 
100 & 10 & - & (a):~$(3/36)\times1$ & - & - & $3/36$\tabularnewline
\hline 
011 & 01 & (a):~$(1/36)\times(\frac{1}{3})$ & (a):~$(3/36)\times(\frac{1}{3})$ & (b):~$(5/36)\times(\frac{2}{15}+\frac{1}{15}+\frac{2}{15})$ & - & $3/36$\tabularnewline
\hline 
\end{tabular}\caption{\label{tab:Weights--for the fundamental rotational transition in HD+}Explanation
of cancellation of QZ shift.\textbf{ Upper table}: Weights $\tilde{\alpha}$
for the frequencies of the transitions connecting to the $J_{z}'=-1$
states of the $N'=1$ rovibrational level in a $N=0\rightarrow N'=1$
transition. The first factor in each entry is, from Fig.~\ref{fig:Spinenergies of rotational transition},
$(2J'+1)/{\cal N}'$, where ${\cal N}'=\sum_{m'}=36.$ The second
factor is obtained from Fig.~\ref{fig:magnetic and el.quadrupole transition diagram},
with the panel letter indicated in the entry. It contains the sum
of the weights of those Zeeman components $J_{z}$ of the given lower
level $FSJ$ that connect to $J_{z}'$. The product of the two factors
is $1/((2|J_{z}'|+1){\cal N}')$, and thus equal for all the considered
states. These equal values enter $f_{t}$ and allow nulling the QZ
shift according to eq.~\ref{eq:Nulling of quadratic ZS}. \textbf{Lower
table}: Same as (a), but for the $J_{z}=0$ states of the $N=0$ level
of the same transition. The sum of weights of the transitions reaching
a particular $FSJ,J_{z}=0$ is shown in the last column. Those values
being all equal allows to null the QZ shift in $f_{t}$, according
to eq.~\ref{eq:Nulling of quadratic ZS}.}
\end{table}
\clearpage{}The same results holds also for the contributions arising
from the other $J_{z}'$ and $J_{z}$ states. Thus, $f_{{\rm t}}^{{\rm pert-free}}$
is free of QZ shift. 
\par\end{center}

The EQ shift is canceled in the summation over all $J_{z}$ and $J_{z}'$.
This also cancels the tensor Stark shift, but not the scalar Stark
shift. It affects all spin states of a given rovibrational level equally.
It is nonzero for all values $N$ and is $v,\,N$ - dependent \cite{Schiller2014a}.
Table~\ref{tab: H2+ and HD+ traceless frequencies with a large number of Zeeman components-1}
gives the values of the Stark shifts for ${\rm H}_{2}^{+}$ and HD\textsuperscript{+},
respectively. The values will not vary much for different choice of
the vibrational levels $v,\,v'$. 

For ${\rm H}_{2}^{+}$, the residual shifts are negligible , below
$1\times10^{-16}$. 

For transitions of HD\textsuperscript{+} involving $N=0$ - levels,
due to their large scalar polarisability, the Stark shifts are not
negligible. Assuming as before $E^{2}\simeq1\,{\rm (kV/m)^{2}}$,
the shift for the fundamental rotational and the fundamental vibrational
transition is approximately $4\times10^{-12}$ and $1\times10^{-13}$,
respectively. We see again the advantage of E2 transitions: the transition
$(0,1)\rightarrow(1,1)$ has a negligible shift and also the smallest
number of transitions (36) among the shown set. This number is, unfortunately,
still rather large.
\begin{center}
\begin{table}[H]
\begin{tabular}{|c||c|c|}
\hline 
\multirow{2}{*}{transition} & \multirow{2}{*}{$n$} & DC Stark shift/$E^{2}$ \tabularnewline
 &  & (Hz (m/kV)$^{2}$)\tabularnewline
\hline 
\hline 
${\rm H}_{2}^{+}:$\textsuperscript{}$\text{E2,\,(0,0)\ensuremath{\to}(1,2)}$ & 10 & -0.0048\tabularnewline
\hline 
${\rm H}_{2}^{+}:$\textsuperscript{}$\text{E2,\,(0,2)\ensuremath{\to}(1,2)}$ & 10 & -0.0046\tabularnewline
\hline 
\hline 
${\rm HD}^{+}:$ $\text{E1,\,(0,0)\ensuremath{\to}(0,1)}$ & 62 & 4.9\tabularnewline
\hline 
${\rm HD}^{+}:$ $\text{E1,\,(0,1)\ensuremath{\to}(0,2)}$ & 142 & -0.00024\tabularnewline
\hline 
${\rm HD}^{+}:$ $\text{E1,\,(0,0)\ensuremath{\to}(1,1)}$ & 62 & 4.9\tabularnewline
\hline 
${\rm HD}^{+}:$ $\text{E1,\,(0,1)\ensuremath{\to}(1,2)}$ & 142 & -0.0092\tabularnewline
\hline 
${\rm HD}^{+}:$ $\text{E1,\,(0,0)\ensuremath{\to}(1,2)}$ & 104 & 4.9\tabularnewline
\hline 
${\rm HD}^{+}:$ $\text{E2,\,(0,1)\ensuremath{\to}(1,1)}$ & 36 & -0.0089\tabularnewline
\hline 
${\rm HD}^{+}:$ $\text{E2,\,(0,2)\ensuremath{\to}(1,2)}$ & 60 & -0.0089\tabularnewline
\hline 
\end{tabular}

\caption{\label{tab: H2+ and HD+ traceless frequencies with a large number of Zeeman components-1}${\rm H}_{2}^{+}$:
Residual systematic shifts contributions to traceless frequencies
$f_{{\rm t}}^{{\rm pert-free}}$ which are sums over all Zeeman components,
according to eq.~(\ref{eq:f_t,tot}). For all cases, the spin structure
contributions cancel, the linear Zeeman shift, the quadratic shift
and the electric quadrupole shifts average to zero.}
\end{table}
\par\end{center}

\section{Largely complete cancellation - optimized method\label{subsec:Largely-complete-cancellation}}

Especially for the case of HD\textsuperscript{+}, the number of transitions
for achieving cancellation of systematics in $f_{{\rm t}}^{\text{pert-free}}$
is large. In this section we present more efficient solutions for
traceless frequencies that have zero LZ, QZ, and EQ shifts, for both
${\rm H_{2}^{+}}$ and ${\rm HD^{+}}$.

\subsection{${\rm H}_{2}^{+}$}

For the $(0,\,0)\rightarrow(1,\,2)$ transition, among the allowed
Zeeman components we select those five having the smallest LZ shifts.
They are marked with red circles and arrows in Fig.~\ref{fig:magnetic and el.quadrupole transition diagram}~(d,e).
We make the ansatz:
\begin{eqnarray}
f_{{\rm t}}^{{\rm opt}}(0,0\rightarrow1,2) & = & \frac{4}{10}\left(\alpha\tilde{f}_{-}\left(J_{z}=\frac{1}{2}\rightarrow J_{z}'=-\frac{1}{2}\right)+(1-\alpha)\tilde{f}_{-}\left(\frac{1}{2}\rightarrow-\frac{3}{2}\right)\right)+\label{eq:f_t for H2+ 5 components}\\
 &  & \frac{6}{10}\left(\beta\tilde{f}_{+}\left(\frac{1}{2}\rightarrow\frac{1}{2}\right)+\gamma\tilde{f}_{+}\left(\frac{1}{2}\rightarrow\frac{3}{2}\right)+(1-\beta-\gamma)\tilde{f}_{+}\left(\frac{1}{2}\rightarrow\frac{5}{2}\right)\right)\,.\nonumber 
\end{eqnarray}
The transitions $\tilde{f}_{-},\,\tilde{f}_{+}$ are indicated in
Fig.~\ref{fig:H2+ energy level scheme}. The spin-structure contributions
cancel for any value of the weights $\alpha,~\beta,~\gamma$. They
are determined by imposing the vanishing of the LZ shift, the QZ shift,
and the EQ shift. The result is shown in Tab.~\ref{tab:traceless frequencies - efficient solution}.
As mentioned above, except for the $\tilde{f}_{+}\left(\frac{1}{2}\rightarrow\frac{5}{2}\right)$
transition, the LZ shifts of the individual Zeeman components in $f_{{\rm t}}^{{\rm opt}}$
are rather large, the largest being 2.5~MHz/G. Therefore the stability
of the magnetic field during the course of the measurement of the
complete set of transitions is a serious issue if ultra-high accuracy
(Hz-level) is aimed for. The table also shows the largest QZ shift,
which in comparison is not important.

For the $(0,\,2)\rightarrow(1,\,2)$ transition, among the allowed
Zeeman components, those having $J_{z}\rightarrow J_{z}'=J_{z}$ exhibit
very small LZ shifts ($<15\,$Hz/G) and small QZ shifts ($<1.2\,{\rm kHz/G^{2}}$).
We choose them for the ansatz:
\begin{eqnarray}
f_{{\rm t}}^{{\rm opt}}(0,2\rightarrow1,2) & = & \frac{4}{10}\left(\alpha f_{-}\left(J_{z}=\frac{1}{2}\rightarrow J_{z}'=\frac{1}{2}\right)+(1-\alpha)f_{-}\left(\frac{3}{2}\rightarrow\frac{3}{2}\right)\right)+\\
 &  & \frac{6}{10}\left(\beta f_{+}\left(\frac{1}{2}\rightarrow\frac{1}{2}\right)+\gamma f_{+}\left(\frac{3}{2}\rightarrow\frac{3}{2}\right)+(1-\beta-\gamma)f_{+}\left(\frac{5}{2}\rightarrow\frac{5}{2}\right)\right)\,.\nonumber 
\end{eqnarray}

The transitions $f_{-},\,f_{+}$ are indicated in Fig.~\ref{fig:H2+ energy level scheme}.
The result of the external-field shift cancellation is reported in
Tab.~\ref{tab:traceless frequencies - efficient solution}. This
is an especially attractive transition, since the residual shift is
negligible. Note that the effect of two spin-structure coefficients
and three systematic shifts are canceled with only 5 transitions.

The $(0,1)\rightarrow(1,1)$ transition requires more components.
The ansatz uses the traceless frequency $f_{{\rm t}}^{{\rm spin}}$
of Eqs.~(\ref{eq:f_c for ortho},\ref{eq:Special case of Jz=00003D1/2 to Jz=00003D-1/2})
(with a particular choice of Zeeman components, leading to LZ shift
of 16.4~kHz/Gauss, a QZ shift of 0.493~kHz/G\textsuperscript{2},
and a EQ shift of 1.76~Hz/(GV/m\textsuperscript{2}) and adds additional
contributions that allow to null LZ, QZ, and EQ shift:

\begin{eqnarray}
f_{{\rm t}}^{{\rm opt}}(0,1\rightarrow1,1) & =f_{{\rm t}}^{{\rm spin}}+ & \frac{4\alpha}{18}\left[f\left(\frac{1}{2},\,\frac{3}{2},\,\frac{3}{2}\rightarrow\frac{1}{2},\,\frac{3}{2},\,\frac{3}{2}\right)-f\left(\frac{1}{2},\,\frac{3}{2},\,\frac{1}{2}\rightarrow\frac{1}{2},\,\frac{3}{2},\frac{1}{2}\right)\right]+\label{eq:f_t_opt for H2+ (0,1)->(1,1)}\\
 &  & \frac{4\beta}{18}\left[f\left(\frac{3}{2},\,\frac{3}{2},\,\frac{3}{2}\rightarrow\frac{3}{2},\,\frac{3}{2},\,\frac{3}{2}\right)-f\left(\frac{3}{2},\,\frac{3}{2},\,\frac{1}{2}\rightarrow\frac{3}{2},\,\frac{3}{2},\,\frac{1}{2}\right)\right]+\nonumber \\
 &  & \frac{6\gamma}{18}\left[f\left(\frac{3}{2},\,\frac{3}{2},\,-\frac{5}{2}\rightarrow\frac{3}{2},\,\frac{3}{2},\,-\frac{5}{2}\right)-f\left(\frac{3}{2},\,\frac{5}{2},\,\frac{5}{2}\rightarrow\frac{3}{2},\,\frac{5}{2},\,\frac{5}{2}\right)\right]\,.\nonumber 
\end{eqnarray}
The numbers in parentheses are $F,J,J_{z}\rightarrow F',J',J_{z}'$.
The resulting solution (see Tab.~\ref{tab:traceless frequencies - efficient solution})
exhibits a large $\gamma$ because of the relatively small LZ shifts
of the two frequencies in the last line as compared to the other.
The largest individual LZ shift is of moderate magnitude, so this
is a viable solution as well. The number of different Zeeman components
is 10 (7 being included in $f_{{\rm t}}^{{\rm spin}}$, see Sec.~\ref{subsec:ft^spin for Ortho-H2+},
and the square parentheses actually containing only 3 additional ones).
With this number, the effect of $5+5=10$ spin structure coefficients
and of three systematic shifts is nulled. Note that the ansatz above
is not unique and is possibly not the most favorable in terms of minimization
of the individual components' shifts.

\subsection{${\rm HD}^{+}$}

For this molecule we construct efficient solutions in a fashion similar
to eq.~(\ref{eq:f_t_opt for H2+ (0,1)->(1,1)}). The main contribution
is the traceless frequency $f_{{\rm t}}^{{\rm spin,LZ}}$ composed
of $J_{z}=0\rightarrow J_{z}'=0$ transitions, so that the LZ shift
is already zero. We add two pairs of Zeeman components of two different
hyperfine transitions ${\rm M}_{1},\,{\rm M}_{2}$. Being pairs, they
to not generate a LZ shift. They allow to null the overall QZ shift
and EQ shift, if their weights are properly chosen. 

As first example, we consider the $(0,0)\rightarrow(1,1)$ transition.
We choose ${\rm M}_{1}:\,FSJ=122\rightarrow F'S'J'=123$ and ${\rm M}_{2}:\,FSJ=011\rightarrow F'S'J'=012$.
The ansatz reads:

\begin{eqnarray}
f_{{\rm t}}^{{\rm opt}}(0,0\rightarrow1,1) & =f_{{\rm t}}^{{\rm spin,LZ}}+ & \frac{7}{36}\left[(\alpha-1)\,f_{M_{1}}\left(J_{z}=0\rightarrow J_{z}'=0\right)\right.+\\
 &  & \frac{1-\alpha}{2}\left.\left(f_{M_{1}}(2\rightarrow3)+f_{M_{1}}(-2\rightarrow-3)\right)\right]+\nonumber \\
 &  & \frac{5}{36}\left[(\beta-1)\,f_{M_{2}}(0\rightarrow0)\right.+\nonumber \\
 &  & \frac{1-\beta}{2}\left.\left(f_{M_{2}}(1\rightarrow2)+f_{M_{2}}(-1\rightarrow-2)\right)\right]\,.\nonumber 
\end{eqnarray}

Here, $f_{{\rm t}}^{{\rm spin,LZ}}$ is the frequency from Tab.~\ref{tab:Examples-of-Jz=00003D0 traceless frequencies-1}.
Note that in each square parenthesis the spin and the LZ contributions
cancel. The individual LZ shifts of the chosen Zeeman components $f_{{\rm M}_{1}},\,f_{{\rm M}_{2}}$
are $\pm0.6\,$kHz/G and $\pm38\,$kHz/G, respectively. The result
of the cancellation is reported in Tab.~\ref{tab:traceless frequencies - efficient solution}.
We see that the total weights of the individual components significantly
reduce their LZ shifts to a small and tolerable value. As discussed,
the scalar Stark shift is substantial for this particular rovibrational
transition, here of order $1\times10^{-13}$. 

A similar procedure can be applied to other rovibrational transitions,
both E1 and E2. The number of Zeeman components increases by 4 in
every case. 

For the E2 transition $(0,1)\rightarrow(1,1)$, we choose for example
${\rm M}_{1}:\,FSJ=122\rightarrow F'S'J'=122$ and ${\rm M}_{2}:\,FSJ=011\rightarrow F'S'J'=011$.
The ansatz reads:

\begin{eqnarray}
f_{{\rm t}}^{{\rm opt}}(0,1\rightarrow1,1) & =f_{{\rm t}}^{{\rm spin,LZ}}+ & \frac{5}{36}\left[(\alpha-1)\,f_{{\rm M}_{1}}\left(J_{z}=0\rightarrow J_{z}'=0\right)\right.+\label{eq:f_t_opt for HD+, E2 (0,1)->(1,1)}\\
 &  & \frac{1-\alpha}{2}\left.\left(f_{{\rm M}_{1}}(1\rightarrow1)+f_{{\rm M}_{1}}(-1\rightarrow-1)\right)\right]+\nonumber \\
 &  & \frac{3}{36}\left[(\beta-1)\,f_{{\rm M}_{2}}(0\rightarrow0)\right.+\nonumber \\
 &  & \frac{1-\beta}{2}\left.\left(f_{{\rm M}_{2}}(1\rightarrow1)+f_{{\rm M}_{2}}(-1\rightarrow-1)\right)\right]\,.\nonumber 
\end{eqnarray}

The solution, shown in Tab.~\ref{tab:traceless frequencies - efficient solution},
indicates that the scalar Stark shift is reduced by approximately
a factor 500, to the $2\times10^{-16}$ level. The closeness of $\beta$
to the value 1 indicates that the inclusion of only the M\textsubscript{1}
pair of transitions already leads to a small EQ shift $(-1.3\,{\rm Hz}\,V_{zz}/({\rm GV/m^{2}}))$.
A second advantage is that the LZ shifts of the individual contributions
in eq.~(\ref{eq:f_t_opt for HD+, E2 (0,1)->(1,1)}) are at most $\pm0.09\,$kHz/G,
so that magnetic field stability is substantially less demanding than
for the E1 transition. 

Additional four rovibrational transitions are shown in the table.
We can state that all HD\textsuperscript{+} transitions shown are
attractive, with the two E2 transitions having $\Delta N=0$ being
particularly so, due to their negligible DC Stark shift.

\begin{sidewaystable}
\begin{tabular}{|c||c|c|c|c|c|}
\hline 
\multirow{2}{*}{transition} & \multirow{2}{*}{$n$} & \multirow{2}{*}{weights} & \multirow{1}{*}{DC Stark shift/$E^{2}$} & ${\rm Max(|QZS_{i}|)}/B^{2}$ & ${\rm Max(|LZS_{i}|)}/B$\tabularnewline
 &  &  & (Hz (m/kV)$^{2}$) & $({\rm kHz/G^{2})}$ & $({\rm kHz/G)}$\tabularnewline
\hline 
\hline 
${\rm H}_{2}^{+}:$\textsuperscript{}$\text{E2,\,(0,0)\ensuremath{\to}(1,2)}$ & 5 & $\left(\frac{1}{2},\,-\frac{11}{3},\,\frac{19}{3}\right)$ & -0.0095 & 48 & 2470\tabularnewline
\hline 
${\rm H}_{2}^{+}:$ $\text{E2,\,(0,2)\ensuremath{\to}(1,2)}$ & 5 & $\left(\frac{1}{2},\,\frac{16}{3},\,-\frac{43}{6}\right)$ & -0.0092 & 3.7 & 0.037\tabularnewline
\hline 
${\rm H}_{2}^{+}:$ $\text{E2,\,(0,1)\ensuremath{\to}(1,1)}$ & 10 & $\left(1.67,\,2.40,\,3.45\times10^{3}\right)$ & -0.0091 & 3.5 & 23\tabularnewline
\hline 
\hline 
${\rm HD}^{+}:$ $\text{E1,\,(0,0)\ensuremath{\to}(0,1)}$ & 14 & $\left(0.337,\,0.715\right)$ & 4.9 & $21$ & $0.79$\tabularnewline
\hline 
${\rm HD}^{+}:$ $\text{E1,\,(0,0)\ensuremath{\to}(1,1)}$ & 14 & $\left(-0.362,\,0.684\right)$ & 4.9 & $23$ & $0.85$\tabularnewline
\hline 
\hline 
${\rm HD}^{+}:$ $\text{E2:\,(0,0)\ensuremath{\to}(1,2)}$ & 16 & $\left(0.792,\,-7.53\right)$ & $4.9$ & $5.1$ & $5.8$\tabularnewline
\hline 
${\rm HD}^{+}:$ $\text{E2,\,(0,1)\ensuremath{\to}(1,1)}$ & 14 & $\left(5.47,\,1.002\right)$ & $-0.0089+3\times10^{-5}\cos^{2}\theta$ & $8.0$ & $0.087$\tabularnewline
\hline 
${\rm HD}^{+}:$ $\text{E2,\,(0,2)\ensuremath{\to}(1,2)}$ & 16 & $\left(26.3,\,3.59\right)$ & $-0.025+0.049\,\cos^{2}\theta$ & $2.6$ & $1.9$\tabularnewline
\hline 
${\rm HD}^{+}:$ $\text{E2,\,(0,2)\ensuremath{\to}(1,0)}$ & 16 & $\left(0.677,\,-1.80\right)$ & $-5.7$ & $4.1$ & $5.4$\tabularnewline
\hline 
\end{tabular}

\caption{\label{tab:traceless frequencies - efficient solution}Residual systematic
shifts contributions to traceless frequencies $f_{{\rm t}}^{{\rm opt}}$
which are produced from a minimized number $n$ of Zeeman components.
For all cases, the spin structure contributions cancel, and the LZ
shift, QZ shift and EQ shift average to zero. The weights in column
3 are $(\alpha,\beta,\gamma)$ for ${\rm H}_{2}^{+}$ and $(\alpha,\beta)$
for ${\rm HD}^{+}$. The numerical values are rounded. The last two
columns list the largest individual effective QZ shift (QZS) and effective
LZ shift (LZS) among all Zeeman components $i$ contributing to $f_{{\rm t}}^{{\rm opt}}$.
Effective here means that the QZS$_{i}$ and LZS$_{i}$ are corrected
for the weight with which the component $i$ contributes.}
\end{sidewaystable}

\subsection{Note}

Karr et al. \cite{Karr2016} proposed the following composite frequency
for nulling of LZ, QZ and EQ shifts of the $(v=0,\,N=0)\rightarrow(v'=2,\,N'=2)$
transition of ${\rm H}_{2}^{+}$, 
\begin{align}
f_{{\rm c}}^{{\rm pert-free}} & =[3\,\tilde{f}_{+}(J_{z}=\pm1/2\rightarrow J_{z}'=\pm5/2)+\\
 & \phantom{=}2\,\tilde{f}_{+}(\pm1/2\rightarrow\pm1/2)+2\,\tilde{f}_{-}(\pm1/2\rightarrow\pm1/2)]/7\,.\nonumber 
\end{align}
A short-hand notation has been used to indicate the average of a Zeeman
component pair. These 6 Zeeman components are marked with brown circles
in Fig.~\ref{fig:magnetic and el.quadrupole transition diagram}~(d,e).
The number is smaller by 4 (2 pairs) compared to the general approach
but larger by 1 compared to the optimized method. Cancellation of
the spin structure contributions was not imposed. Indeed, $f_{c}^{{\rm pert-free}}=f^{{\rm spin-avg}}+[10(N'c_{e}'/2)+4(-(N'+1)c_{e}'/2)]/14=f^{{\rm spin-avg}}+2c_{e}'/7$.

\section{Discussion and Conclusions}

We have discussed approaches to null the effect of the perturbation
energy shifts arising in the lower and upper levels of quantum systems
by considering composite transition frequencies. A fundamental property
of certain hamiltonians, the tracelessness, and the structure of the
perturbation expressions, must be used as guiding principles for identifying
a set of contributing transition frequencies and their corresponding
weights. It is remarkable that the number of transitions required
to cancel the effect of the spin hamiltonian coefficients can be smaller
than the latter number, e.g. for the $(0,1)\rightarrow(1,1)$ transition
in HD\textsuperscript{+} the numbers are 10 transitions (in $f_{{\rm t}}^{{\rm spin}}$)
vs. $9+9=18$ coefficients. The reason is the algebraic structure
of the spin hamiltonian.

We showed that by an extended combination $(f_{{\rm t}}^{{\rm pert-free}})$
of Zeeman-resolved transition frequencies not only the spin structure,
but also the four contributions: linear Zeeman, quadratic Zeeman,
electric quadrupole, and tensor Stark shift can be canceled simultaneously.
The scalar Stark shift, for which a cancellation cannot occur if only
a single rovibrational transition is addressed, is at the $10^{-16}$
level for particular suitably chosen rovibrational transitions, for
both ${\rm H}_{2}^{+}$ and HD\textsuperscript{+}.

We also showed that ``economic'' combinations of transition frequencies,
$f_{{\rm t}}^{{\rm opt}}$, can also provide both spin structure and
external-field-shifts cancellations.

A more limited case is spin-structure-shift cancellation and nulling
of the LZ shift ($f_{{\rm t}}^{{\rm spin,LZ}}$), which can be achieved
simultaneously by specifically interrogating only the $J_{z}=0\rightarrow J_{z}'=0$
components or the $J_{z}=-1/2\rightarrow J_{z}'=-1/2$ and $J_{z}=1/2\rightarrow J_{z}'=1/2$
component pairs in case of systems with integer or half-integer angular
momentum, respectively. The QZ, the EQ and Stark shifts then remain
present. 

We have treated the particular systems ${\rm H_{2}^{+},\,HD^{+}}$
in detail because of a specific metrological application of these
molecules, but the treatment is applicable to any quantum system with
spin structure. The potential utility of the proposed approach is
clearly highest when applied to rotational transitions, where the
contributions of the spin energies and of the systematic shifts to
the transition frequency are, in fractional terms, approximately two
orders larger than for vibrational frequencies. The metrological interest
of rotational transitions of the molecular hydrogen ions is as follows.
A rotational frequency is closely proportional to $R_{\infty}m_{e}/\mu$,
where $\mu$ is the reduced nuclear mass. Vibrational transition frequencies
between levels having small $v,\,v'$ are instead closely proportional
to $R_{\infty}\sqrt{m_{e}/\mu}$. Therefore, the measurement of one
experimental spin-averaged rotational frequency and one vibrational
frequency and comparison with ab initio predictions can lead to an
\textit{independent} determination of $R_{\infty}$ and $m_{e}/\mu$. 

For ${\rm H}_{2}^{+}$ the implementation of the proposed method appears
promising. If the rovibrational level $(v=0,\,N=0)$ or $(v=0,\,N=2)$
is chosen as initial level, \textit{just two} spin-resolved transitions
need to be measured in order to cancel the spin structure of a vibrational
transition. Moreover, these levels are very suitable for spectroscopy
because they contain only a single or two spin states, respectively,
so that their preparation is simplified \cite{Schiller2017}. Third,
some of the Zeeman components exhibit very small systematic effects
already for individual Zeeman components \cite{Schiller2014,Karr2016}.
Two particular combinations of 4 Zeeman transitions cancel both the
LZ and the QZ shift, leaving as dominant shift the EQ shift of approximately
$1\times10^{-14}$ in the more advantageous of the two. If in addition
the EQ shift is to be removed, 5 transitions are to be measured, leading
to a residual (Stark) shift of $1\times10^{-16}$. 

For ${\rm HD}^{+}$ the implementation effort is, at first sight,
larger, because of the more complex spin structure: for example, 10
Zeeman components need to be measured in order to extract the spin-averaged
frequency of the fundamental rotational or vibrational transition.
However, with proper choice of the Zeeman components, the LZ shifts
are canceled with the same number of transitions. We have proposed
to use a E2 transition, and again 10 Zeeman components, so as to reduce
the other systematic shifts to the low-$10^{-13}$ level. 

The shifts can be completely nulled by increasing to 14 or 16 the
number of Zeeman components to be measured. Then, only the scalar
Stark shift remains, of order $10^{-16}$ for the E2 transitions. 

This level is comparable to that arising from black-body radiation
at 300~K, since the r.m.s. electric field assumed here is similar
to that produced by the radiation at this temperature.

The above discussion has treated explicitly the cases of small rotational
angular momenta $N=0,\,1,\,2$, so it included the fundamental rotational
transitions, $(v=0,\,N=0)\rightarrow(0,\,1)$ for ${\rm HD}^{+}$
and $(v=0,\,N=0)\rightarrow(0,\,2)$ for ${\rm H}_{2}^{+}$, but is
not limited to these. 

Rotational transition frequencies can also be determined indirectly,
via subtraction of vibrational transition frequencies. This can be
an effective approach in the case of ${\rm H}_{2}^{+}$, for which
a suitable source for driving the E2 fundamental rotational transition
at 5.3~THz may not be available, but which can also be obtained as
$f(0,\,0\rightarrow v',\,2)-f(0,\,2\rightarrow v',\,2)$. In order
to achieve the spin-structure independence of this computed rotational
frequency, the present approach would be applied to the two contributing
vibrational transitions. A first analysis indicates that the smallest
shift for a ($N=0\rightarrow N'=1,2$) computed rotational transition
frequency appears to be obtainable by combining an E1 and an E2 vibrational
transition of HD\textsuperscript{+}, whose scalar Stark shifts would
cancel. The shift, $10^{-14}$ and lower, would be smaller than if
the $(0,0)\rightarrow(0,1)$ rotational transition is measured directly. 

The vibrational transitions discussed here were the fundamental ones,
$v=0\rightarrow v'=1$. The results will be similar for the overtones
$v=0\rightarrow v'>1$. The fractional residual shifts will decrease
with increasing $v'$ since the shifts themselves do not scale with
$v'$.

Finally, we emphasize that the concept of traceless frequency is valuable
and can be applied irrespective of future progress in the \textit{ab
initio} theory of the spin hamiltonian. Of course such progress \textit{is}
important, allowing experimental tests of the higher-order-in-$\alpha$-corrections
to the spin hamiltonian coefficients by probing \textit{individual}
spin components. Concerning experimental feasibility of the present
proposal we remark that already in Ref.~\cite{Bressel2012} a large
number (11) of hyperfine components of a particular rovibrational
transition were measured. Thus, the experimental effort necessary
to determine traceless frequencies appears manageable.

\smallskip{}

\begin{acknowledgments}
V.~I.~K. acknowledges support from the Russian Science Foundation
under Grant No. 18-12-00128.

Corresponding author, step.schiller@hhu.de
\end{acknowledgments}

\bibliographystyle{elsarticle-num}

\cleardoublepage{}
\end{document}